\documentclass[a4paper,11pt]{article}

\usepackage{jheppub} % for details on the use of the package, please

\usepackage{graphicx}% Include figure files
\usepackage{color}
\usepackage{mathtools}
\usepackage[left]{lineno}
\usepackage{subfig}

\title{Mass hierarchy discrimination with atmospheric neutrinos in large volume ice/water Cherenkov detectors}

\author[a,1]{D.Franco,\note{Corresponding author: dfranco@in2p3.fr}}
\author[b]{C.Jollet,}
\author[a]{A.Kouchner,}
\author[a,c]{V.Kulikovskiy,}
\author[b,2]{A.Meregaglia,\note{Corresponding author: amerega@in2p3.fr}}
\author[a]{S.Perasso,}
\author[b]{T.Pradier,}
\author[a]{A.Tonazzo}
\author[a]{and V.Van Elewyck}

\affiliation[a]{APC, Universit\'e Paris Diderot, CNRS/IN2P3, CEA/IRFU, Observatoire de Paris, Sorbonne Paris Cit\'e, 75205 Paris, France}
\affiliation[b]{IPHC, Universit\'e de Strasbourg, CNRS/IN2P3, 67037 Strasbourg, France}
\affiliation[c]{INFN - Sezione di Genova, 16146 Genova, Italy}

\date{\today}
\abstract{
Large mass ice/water Cherenkov experiments, optimized to detect low energy (1--20~GeV) atmospheric neutrinos,  have the potential to discriminate between normal and inverted neutrino mass hierarchies.  The sensitivity depends on several model and detector parameters, such as the neutrino flux profile and normalization, the Earth density profile, the oscillation parameter uncertainties, and the detector effective mass and  resolution. \\
A proper evaluation of the mass hierarchy  discrimination power requires a robust statistical approach. In this work, the Toy Monte Carlo, based on an extended unbinned likelihood ratio test statistic, was used. The  effect of each model and detector parameter, as well as the required detector exposure, was then studied.  While uncertainties on the Earth density and atmospheric neutrino flux profiles were found to have a minor impact on the mass hierarchy  discrimination, the flux normalization, as well as some of  the oscillation parameter ($\Delta m^2_{31}$, $\theta_{13}$, $\theta_{23}$, and $\delta_{CP}$) uncertainties and correlations resulted critical. Finally, the minimum required detector exposure, the optimization of the  low energy threshold, and the detector resolutions were also investigated.
}

\begin{document} 

%\maketitle must follow title, authors, abstract, \pacs, and \keywords
\maketitle
\flushbottom

% body of paper here - Use proper section commands
% References should be done using the \cite, \ref, and \label commands

\section{Introduction}
The field of neutrino physics has witnessed important theoretical and experimental progress in the past decade. A variety of experiments with solar, atmospheric, reactor and accelerator neutrinos   spanning energies from the MeV up to tens of GeV have provided compelling evidence that the known flavour eigenstates ($\nu_e, \nu_\mu, \nu_\tau$) mix, implying the existence of non-zero neutrino masses; see e.g. the review by Nakamura and Petcov in ref.~\cite{Beringer:1900zz}.

 In the standard $3\nu$ scheme, the mixing matrix which relates the neutrino flavour eigenstates to the mass eigenstates ($\nu_1, \nu_2, \nu_3$) can be parameterized in terms of 3 mixing angles $\theta_{12}$, $\theta_{13}$ and $\theta_{23}$,  and a Dirac CP-violation phase $\delta_{CP}$. Oscillation experiments are not sensitive to the absolute value of neutrino masses but they provide measurements of the squared-mass splittings $\Delta m^2_{ij} = m_i^2 - m_j^2$ ($i,j=1,2,3$).
In this scheme, there are only two independent squared-mass differences: one ($\delta m^2 \simeq 7.5\times 10^{-5}\, \mathrm{eV}^2$) is %conventionally 
associated to the mass splitting arising from solar (and reactor) observations, while the other ($\Delta m^2 \simeq  2.5\times 10^{-3}\, \mathrm{eV}^2$) is extracted from the atmospheric neutrino sector.

The values of these mass splittings, as well as those of the three mixing angles, are now extracted from global fits of available data with a reasonable precision~\cite{Tortola:2012te,Fogli:2012ua,GonzalezGarcia:2012sz}.
%A consequence is that the ordering of neutrino mass eigenstates is not univocally determined so far; two solutions remain possible, usually dubbed as normal hierarchy (NH: $m_1 < m_2 < m_3$, with $\Delta m^2_{21} \approx \delta m^2$ and $\Delta m^2_{32} \simeq \Delta m^2_{31} \approx \Delta m^2$) and inverted hierarchy (IH: $m_3 < m_1 < m_2$, with $\Delta m^2_{21} \approx \delta m^2$ and $\Delta m^2_{23} \simeq \Delta m^2_{13} \approx \Delta m^2$).\\
%The values of all mixing angles and squared-mass differences in the $3\nu$ oscillation scheme can now be extracted from global fits of available data with a reasonable precision~\cite{Tortola:2012te,Fogli:2012ua,GonzalezGarcia:2012sz}. %, meaning that the measurement of neutrino oscillation parameters is entering a precision era. 
Last but not least, the recent observation of  $\overline{\nu}_e$ disappearance in several short-baseline reactor experiments~\cite{Abe:2011fz,Ahn:2012nd,Xing:2012ej} has provided the first high significance measurement of the mixing angle $\theta_{13}$ which drives the $\nu_\mu - \nu_e$ transition amplitude. The relatively large value of this parameter, $\sin^2(2\theta_{13})\simeq 0.1$, is an asset for the subsequent searches for the remaining major unknowns of the model, in particular the possible presence of a CP-violating phase in the neutrino sector, and the neutrino mass hierarchy (NMH). 
The ordering of neutrino mass eigenstates is indeed not univocally determined so far; two solutions remain possible, usually dubbed as normal hierarchy (NH: $m_1 < m_2 < m_3$, with $\Delta m^2_{21} \equiv \delta m^2$ and $\Delta m^2_{32} \simeq \Delta m^2_{31} \equiv \Delta m^2$) and inverted hierarchy (IH: $m_3 < m_1 < m_2$, with $\Delta m^2_{21} \equiv \delta m^2$ and $\Delta m^2_{23} \simeq \Delta m^2_{13} \equiv \Delta m^2$).%has yielded the important  
%all these parameters are known with a precision better than 15\%, the largest remaining uncertainty being currently on $\sin^2{\theta_{23}}$.\\
%(d\'etailler ???? octant, non-maximal mixing 23). (citer importance des nu atmospheriques dans ces rŽsultats). 

The question of NMH discrimination is on the agenda of most current and next-generation neutrino experiments in the GeV energy domain\footnote{Note that the possibility of using reactor neutrinos to measure NMH~\cite{Petcov:2001sy} has also been reconsidered in view of the new perspectives opened by the large value of $\theta_{13}$~\cite{Qian:2012xh,Ciuffoli:2012bs,Ghoshal:2012ju}.}. 
Its determination at high confidence level (5$\,\sigma$ or more), however, remains challenging and could take as long as 15 to 20 years even within optimistic estimates~\cite{Itow:2001ee,Ayres:2004js,Schwetz:2006ti,LAGUNA,Samanta:2006sj,Ghosh:2012px,Abe:2011ts,Agarwalla:2012zu,Barger:2012fx}. The reach of accelerator experiments is indeed strongly dependent on the still unknown value of $\delta_{CP}$~\cite{Huber:2009cw}, while atmospheric neutrino experiments suffer from the large uncertainties in the fluxes and from the relatively low statistics. The latter experiments have nonetheless the advantage of probing $\nu_\mu - \nu_e$ oscillations over a wide range of propagation distances (or baselines) and neutrino energies, accessing thereby the richness of matter effects that arise in the propagation of neutrinos through the Earth~\cite{Bernabeu:2001xn} (see also collective references in ref.~\cite{Ghosh:2012px} for example).

In this realm, it has been pointed out recently that large water/ice Cherenkov detectors, such as the ones currently used for neutrino astronomy, could also be competitive in providing a measurement of the NMH based on the study of GeV atmospheric neutrinos~\cite{Mena:2008rh, Akhmedov:2012ah,Agarwalla:2012uj}. These instruments were primarily designed for the detection of extra-terrestrial neutrinos in the TeV to PeV energy range, but they also measure  atmospheric neutrinos.
They consist in 3-dimensional arrays of photomultipliers (PMTs) that detect the Cherenkov light originating from the products of the neutrino interaction in and around the instrumented volume. The arrival direction of the parent neutrino and its energy can be inferred from the timing, position and amplitude of the hits recorded by the PMTs. The directional accuracy is best for $\nu_\mu$-induced muon tracks, which can be reconstructed even if they are not fully contained inside the detector volume. The detection of contained showers is also possible, with a relatively accurate estimation of the neutrino energy based on the measurement of the total amount of light deposited in the detector. The optimization of such instruments for a given neutrino energy range is therefore a trade-off between the total target volume and the density of photosensors. Although they do not have charge identification capabilities, these detectors could observe the imprint of the NMH in the pattern of events of a given flavor detected at different energies and baselines, provided their angular and energy accuracies are sufficient.

The largest Cherenkov neutrino telescopes currently operating are IceCube~\cite{Halzen:2010yj}, with an instrumented volume of 1 km$^3$ of South Polar ice, and ANTARES~\cite{Collaboration:2011nsa}, a smaller array located in the Mediterranean Sea off the coast of France, which is also a prototype for the future multi-km$^3$ neutrino Cherenkov detector to be built in the Mediterranean, KM3NeT~\cite{km3net,km3net2}. While both detectors reach their full capabilities for neutrinos of energy of the TeV and beyond, significant effort has been made to improve their performances at lower energies as well. ANTARES has recently proven its capability to detect atmospheric muon neutrinos with energies down to about 20 GeV, providing the first measurement of neutrino oscillation parameters obtained with a large-volume neutrino telescope~\cite{AdrianMartinez:2012ph}. IceCube has been complemented by a denser infill dubbed as DeepCore, allowing to lower its energy threshold down to about 10 GeV~\cite{deepcore}.%~\cite{Sullivan:2012hf}PARLER DU VETO+ resultat oscillations.

A study of neutrino hierarchy in the atmospheric sector, at energies in the range 1 -- 50 GeV, would however require a new generation of detectors with a different layout, and in particular a smaller but denser array of photosensors. Muons with energies of a few GeV indeed travel distances of the order of ten meters. For this reason the distance between the optical modules has to be of the order of a few meters, much less than in currently operating detectors. %The closer the PMTs, the higher the 
A denser array provides better energy and angular resolution; however, for the same total number of PMTs, the fiducial volume gets reduced. Therefore a compromise for an optimal spacing of the optical modules has to be found. Such optimization studies are underway for both detector sites.

At the South Pole, the PINGU project~\cite{Koskinen:2011zz,DeYoung:2011} studies the possibility to deploy an even denser array of about 20 lines of photosensors within DeepCore, to benefit from the vetoing capabilities of the surrounding detectors. For the Mediterranean site, the ORCA project~\cite{ORCAstatusreport} proposes to build a similar, dense array of about 50 lines as a first phase of construction of the KM3NeT detector.%, for which part of the funding is already available. %Optimization studies are ongoing within both collaborations,`
At this stage it is not clear yet which performances can be achieved in terms of effective volume, angular and energy reconstruction, and flavor identification, which are the main ingredients for the NMH discrimination. As no official numbers exist so far for PINGU or ORCA effective volumes, in this study the assumption made in ref.~\cite{Akhmedov:2012ah} for the effective mass $\mathrm{M_{eff}}$ has been followed, namely:
\begin{equation}
\label{eq:mass}
\mathrm{M_{eff}} \propto (\log_{10}E_\nu)^{1.8}.
\end{equation}
Note that the dependence of $\mathrm{M_{eff}}$ on the neutrino energy $E_\nu$ could vary according to the actual performances of the detector, in particular at low energies. Throughout this paper, the effective mass normalization has been anchored at 40 GeV, meaning that all results will be presented in terms of {\it effective exposure} defined in units of megaton year (Mt $\times$ year) at 40 GeV. With this convention, the effective volume quoted for PINGU in ref.~\cite{Akhmedov:2012ah} corresponds to 34.1 Mt.

The intrinsic detector capabilities are not the only factor that could spoil the measurement of NMH. Uncertainties in the incident atmospheric neutrino flux will affect the initial $\nu_\mu/\nu_e$ ratio, while those related to the neutrino oscillation parameters and the Earth density profile can further modify the flavor content of oscillated neutrinos that are observed at the detector level. Last but not least, at such low energies the kinematics of neutrino interaction also induces an intrinsic uncertainty on the incident neutrino direction, even for $\mu$-like events. All these effects have to be carefully studied in order to evaluate their potential impact on the NMH measurement. In addition, an appropriate statistical method has to be used to properly estimate the confidence level with which one can hope to discriminate between NH and IH.

In this paper a method to evaluate the discrimination power of large Cherenkov neutrino detectors for NMH is presented, based on a Toy Monte Carlo (MC) approach and, as test statistic, an extended unbinned log-likelihood ratio. The details of the statistical method are presented in section~\ref{sec:method}, while section~\ref{sec:MC} describes the MC chain and its main ingredients. Details on the choice of the reference oscillation parameters can be found in section~\ref{Introsys}. Results in terms of NMH discrimination power as a function of the effective exposure of the detectors are given in section~\ref{sec:exposure}. To further illustrate the method, studies are conducted to quantify the impact of the uncertainties listed hereabove on the discrimination power of the experiments. Sections~\ref{sec:fluxsys},~\ref{sec:profsys} and~\ref{sec:oscilsys} discuss the systematics related to the atmospheric neutrino flux, Earth density profile and oscillation parameters respectively. Preliminary hints on the potential impact of the detector energy and angular resolution are also provided in section~\ref{sec:resolution}. Conclusions are drawn in section~\ref{sec:conclusion}.

% The fiducial volume increases with the energy of the neutrinos to be detected, since high energy muons travel longer distances and therefore the volume around the instrumented one in which the interaction can take place and still be detected is larger.\\

%Two detectors based on this technique could be used in particular in the near future for the mass hierarchy determination, namely PINGU~\cite{Koskinen:2011zz} and ORCA.\\

% talk Mezzetto:
%T2K/NOVA most optimistic scenario for mass hierarchy Huber et al JHEP 0911:044
%hierarchy from reactors: 50 km baseline maximises the effect, small wiggles --> need excellent energy resolution (goal 3%/sqrt(E)), good linearity + 10 x Kamland statistics
% good competitor: Daya Bay -II with new detector at 60 km, significant reduction of uncertainties on theta12, dm12, 
% possible construction 2016-2020 ?
% with atmospherics:
%INO: magnetic field --> discriminate nu/antinu, some power in hierarchy discrimination
%2 sigma in 5 years by 2022
% HyperKamiokande: need to measure subleading effects in nu_e (because nu£_mu are not contained in energy range of interest: few GeV --> bad energy resolutions)

%1210.3651 : Gaussian statistics is not the most appropriate because the herarchy can only take 2 dsicrete values: +/- 1

%QUESTIONS
% - resonant matter effects included in GLOBeS ?
% - 

%\section{Detectors}
%\input{Detectors}

\section{Statistical method}
\label{sec:method}

\begin{figure}[t]
\begin{center}
 \begin{minipage}{0.6\linewidth}
\includegraphics[width=\columnwidth]{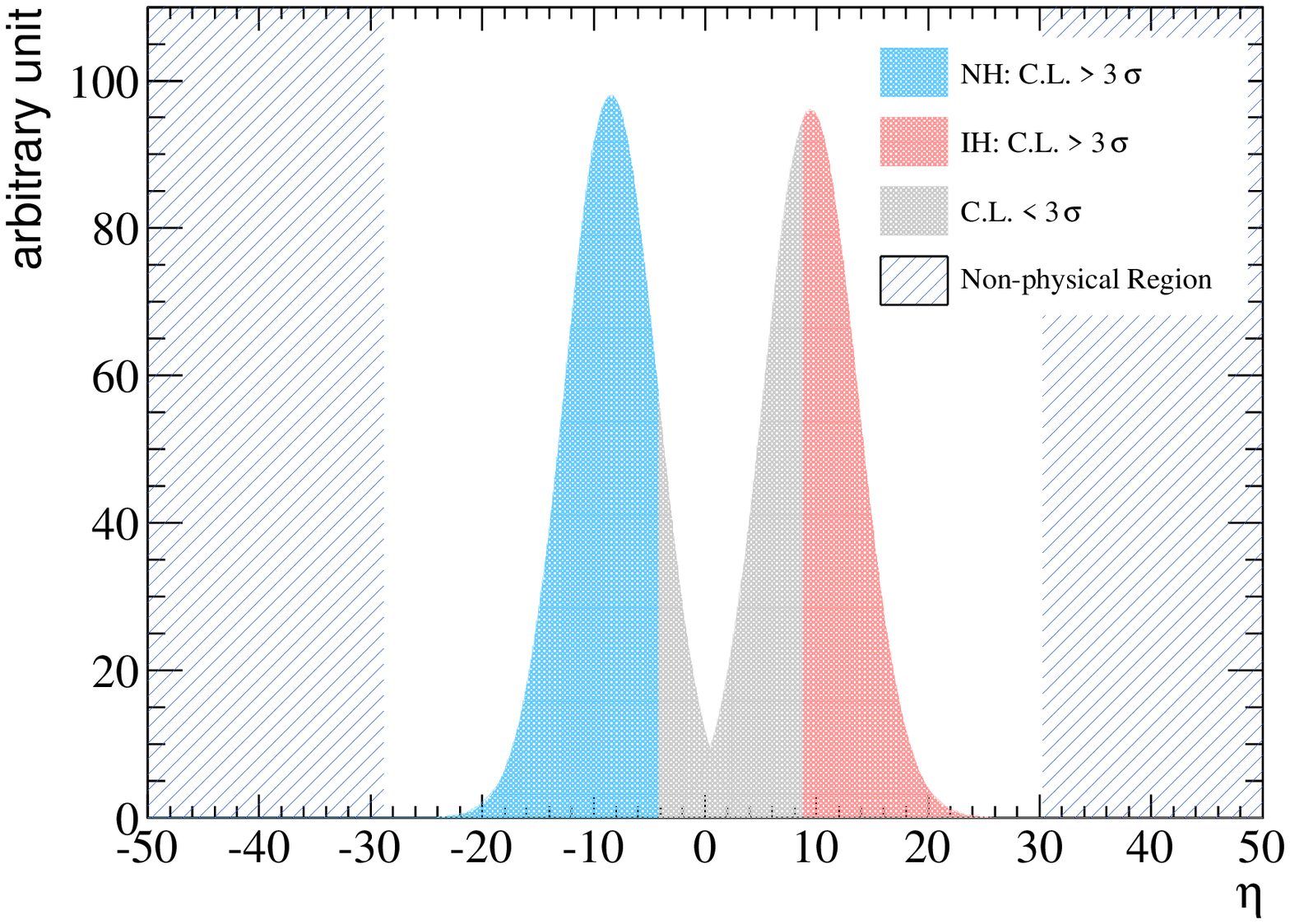}
\end{minipage}
\hspace{0.5cm}
 \begin{minipage}{0.6\linewidth}
\includegraphics[width=\columnwidth]{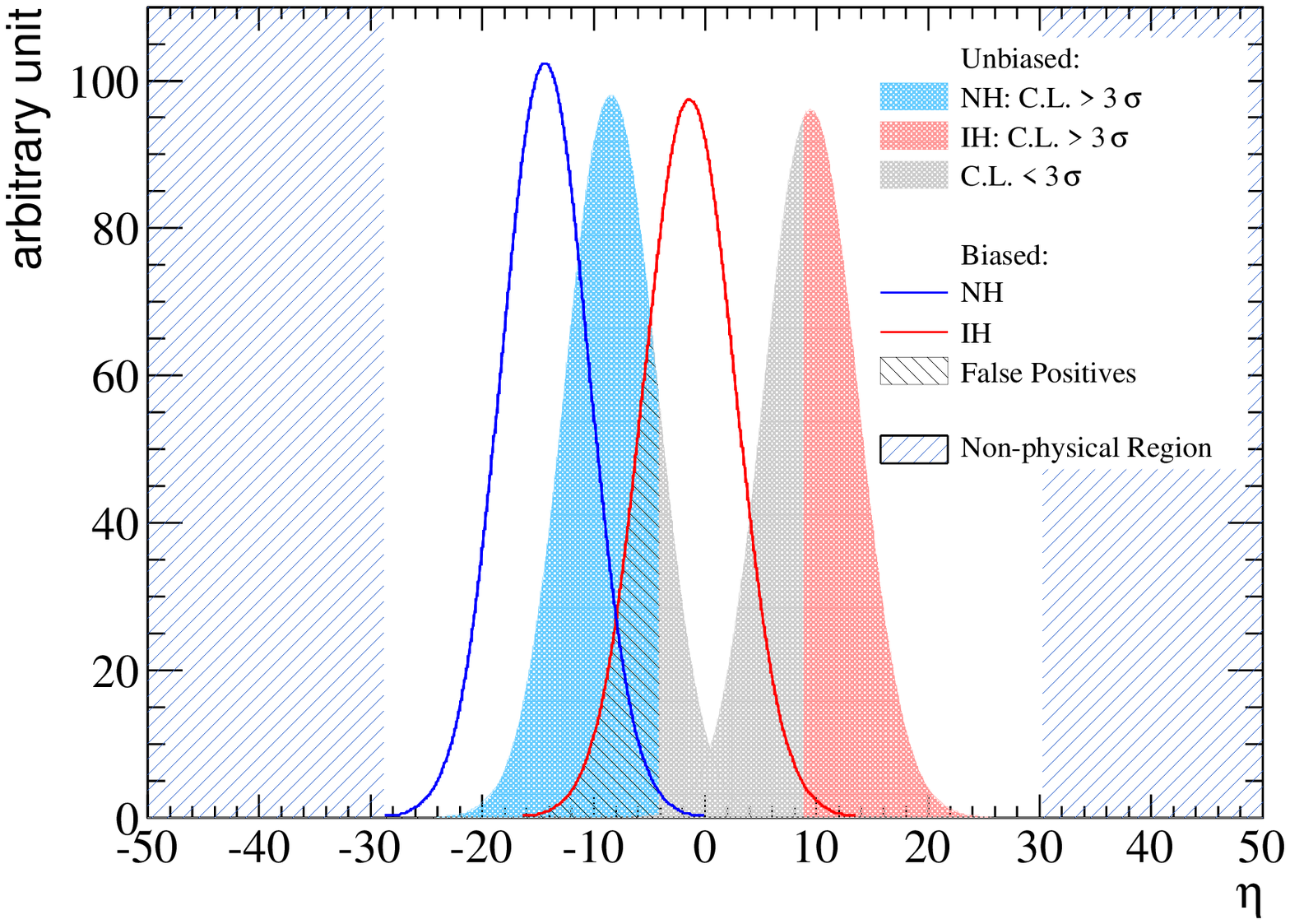}
\end{minipage}
\caption{Example (top) of test statistic $\eta$ distributions. The blue (red) shaded region corresponds to the range of $\eta$ where NH (IH) is identified at 3$\,\sigma$ C.L.. In grey the region for which no mass hierarchy determination can be achieved is shown. The hatched blue areas indicate regions of unphysical results. \\
Comparison (bottom)  between unbiased (shaded areas), as in the top plot, and biased (solid lines) $\eta$ distributions. The hatched black region corresponds to the false positives.}
\label{fig:pval}
\end{center}
\end{figure}
The Toy Monte Carlo approach is a flexible tool to test the NMH discrimination power of future ice/water detectors. It allows to investigate the discrimination power dependence on oscillation parameter uncertainties, neutrino flux and Earth profile models,  as well as detector exposure and systematic effects.\\
The work scheme requires first to fix the \textit{true hypothesis} under investigation, NH or IH, for a given set of parameters and models. On the basis of the true hypothesis, 1000 test experiments are generated,  event--by--event, with the corresponding event statistics. Each  test experiment is then compared with the \textit{model hypothesis},  by evaluating the following extended unbinned likelihood:
 \begin{equation}
 \label{eq:likelihood}
L_j = \frac {(e^{-\mu_j} \mu_j^n )}{n!} \times \prod_{i=1}^n \mathrm{pdf}_j(E_i,\theta_i)
\end{equation}
where $\mu_j$ is the expected number of events with $j=\{$NH, IH$\}$, $n$ the number of observed events, and pdf$_j(E_i, \theta_i)$ the probability of observing the $i^{th}$ event with energy  $E_i$ and zenith angle $\theta_i$. The probability density function pdf$_j(E, \theta)$ represents the model hypothesis, and it is produced with high--statistics Monte Carlo simulation (1000 times the expected statistics).

The test statistic $\eta$ used to evaluate the mass hierarchy discrimination power is the logarithm of the  likelihood ratio  between IH and NH:
\begin{eqnarray}
\eta & = & \log (L_{\mathrm{IH}}/L_{\mathrm{NH}}) \nonumber \\
 & =  & - (\mu_\mathrm{IH} - \mu_\mathrm{NH}) + n \log (\mu_\mathrm{IH}/\mu_\mathrm{NH}) \nonumber \\
 & & +  \sum_i \log(\mathrm{pdf}_\mathrm{IH}(E_i, \theta_i)/\mathrm{pdf}_\mathrm{NH}(E_i, \theta_i)).
\end{eqnarray}
The $\eta$ distribution is produced for each true hypothesis, NH and IH, each entry corresponding to a test experiment. In order to attenuate the statistical fluctuations, each distribution is then fitted with a Gaussian function. The Gaussianity of the so-produced distributions was demonstrated with dedicated high--statistics tests.\\
Finally, the probability (p-value) to achieve the confidence level $\alpha$ (in this work equivalent to 3 or 5$\,\sigma$) in the hierarchy discrimination, was defined as the fraction of test experiments yielding a value of $\eta$ satisfying:
\begin{equation}
\frac{N_{t}(\eta)} {N_\mathrm{NH}(\eta) + N_\mathrm{IH}(\eta)} > \alpha
% \frac{N_{NH}(\eta)} {N_{NH}(\eta) + N_{IH}(\eta)} > \alpha \, \, \, \, \mathrm{OR} \, \, \, \, \frac{N_{IH}(\eta)} {N_{NH}(\eta) + N_{IH}(\eta)} > \alpha
\end{equation}
for either $t$ =  NH or $t$ = IH, where $N_t(\eta)$ is the number of experiments corresponding to the true hypothesis $t$.

For illustrative purposes, the  $\eta$ distributions obtained for a standard case are shown in figure \ref{fig:pval}. The blue (red) shaded region corresponds to the fraction of areas where NH (IH) is identified at more than 3$\,\sigma$ C.L.. On the contrary, there is no sensitivity to the NMH discrimination in the grey shaded region of overlap of the two distributions. Furthermore,  it was assumed that the testing model is strongly affected by systematics (and hence the result unphysical), if the measurement is  more than 5$\,\sigma$ away from any of the Gaussian mean values, which corresponds to the hatched blue vertical bands in figure~\ref{fig:pval}.

In order to investigate the impact of the model parameters and their uncertainties, on the hierarchy discrimination, biases are introduced in the true hypothesis, maintaining unvaried the model hypothesis.  The bias can lead to false positives, when one of the two hierarchies is wrongly recognized. An example is shown in figure  \ref{fig:pval}, where the value of $\Delta$m$^2_{31}$ is reduced by 1$\,\sigma$ in the true hypothesis.  In order to quantify the fraction of false positives, $\eta$ distributions are evaluated with respect to both the unbiased and biased hypotheses. As  net effect, the biased hypothesis distributions are shifted with respect to the unbiased ones. Small variations in the Gaussian sigmas are also observed. The fraction of false positives is evaluated after subtracting the unphysical region.\\ 
A general description of the adopted statistical method can be found in ref.~\cite{Lyons:2012}.  Another statistical approach can be found in ref.~\cite{Qian:2012zn}.\\

In the next sections, p--value is quoted arbitrarily at the confidence level (3 or 5$\,\sigma$) that best highlights its variations.

\section{The Toy Monte Carlo approach}
\label{sec:MC}
In order to evaluate the mass hierarchy sensitivity, a specific Monte Carlo simulation was set up.\\
The full MC chain is made up of different steps %and the basic ingredients needed are:
 that rely on the following ingredients: the neutrino fluxes, the oscillation probabilities, the Earth density profile and the neutrino cross sections. In addition, detector-specific information on the event reconstruction has to be implemented for a realistic sensitivity evaluation.
 
As far as the neutrino flux is concerned, several models are available. As a base option the Honda model~\cite{Honda:1995} was used. A detailed discussion on the fluxes and a comparison between the different models can be found in section~\ref{sec:fluxsys}.

To compute the neutrino oscillation probability when propagating through the Earth, the GLoBES software tool~\cite{Huber:2007ji,Huber:2004ka} was used.\\
Given a set of oscillation parameters, the code provides the oscillation probability either in vacuum or when traveling through matter. An option to calculate oscillation probabilities when neutrinos traverse the Earth is already available. The distance travelled through the Earth (baseline) is split into a number of given steps, each with a mean constant density computed using the Preliminary Reference Earth Model (PREM)~\cite{Dziewonski,Stacey}, according to the distance from the Earth's core. In the simulation 1000 steps were fixed for each baseline.
The baseline depends on the zenith angle $\theta$ of the incoming neutrinos, therefore oscillation probabilities were generated for fixed values of $\cos \theta$ ranging from -1 to 0 with a step of 0.02.

To compute the number of expected events, the charged current cross sections for $\nu_\mu$ and $\bar{\nu}_\mu$ already provided in the GLoBES simulation toolkit have been used~\cite{Paschos:2001np,Messier:1999kj}.

Two-dimensional matrices (energy versus $\cos \theta$) containing the number of expected $\nu_{\mu}$ + $\bar{\nu}_{\mu}$ events were computed, both for NH and IH, using the selected atmospheric neutrino flux as a function of $\cos \theta$, the oscillation probability for several sets of oscillation parameters and assuming a fixed effective mass of 1 Mt.\\
Note that no discrimination was assumed between neutrinos and antineutrinos and that the following components were considered: $\nu_\mu \to \nu_\mu$, $\nu_e \to \nu_\mu$,  $\bar{\nu}_{\mu} \to  \bar{\nu}_{\mu}$ and $\bar{\nu}_e \to  \bar{\nu}_\mu$.\\
In addition, downgoing neutrinos were not taken into account in the MC.
\begin{figure}[t]
\begin{center}
\begin{minipage}{0.6\linewidth}
\includegraphics[width=\columnwidth]{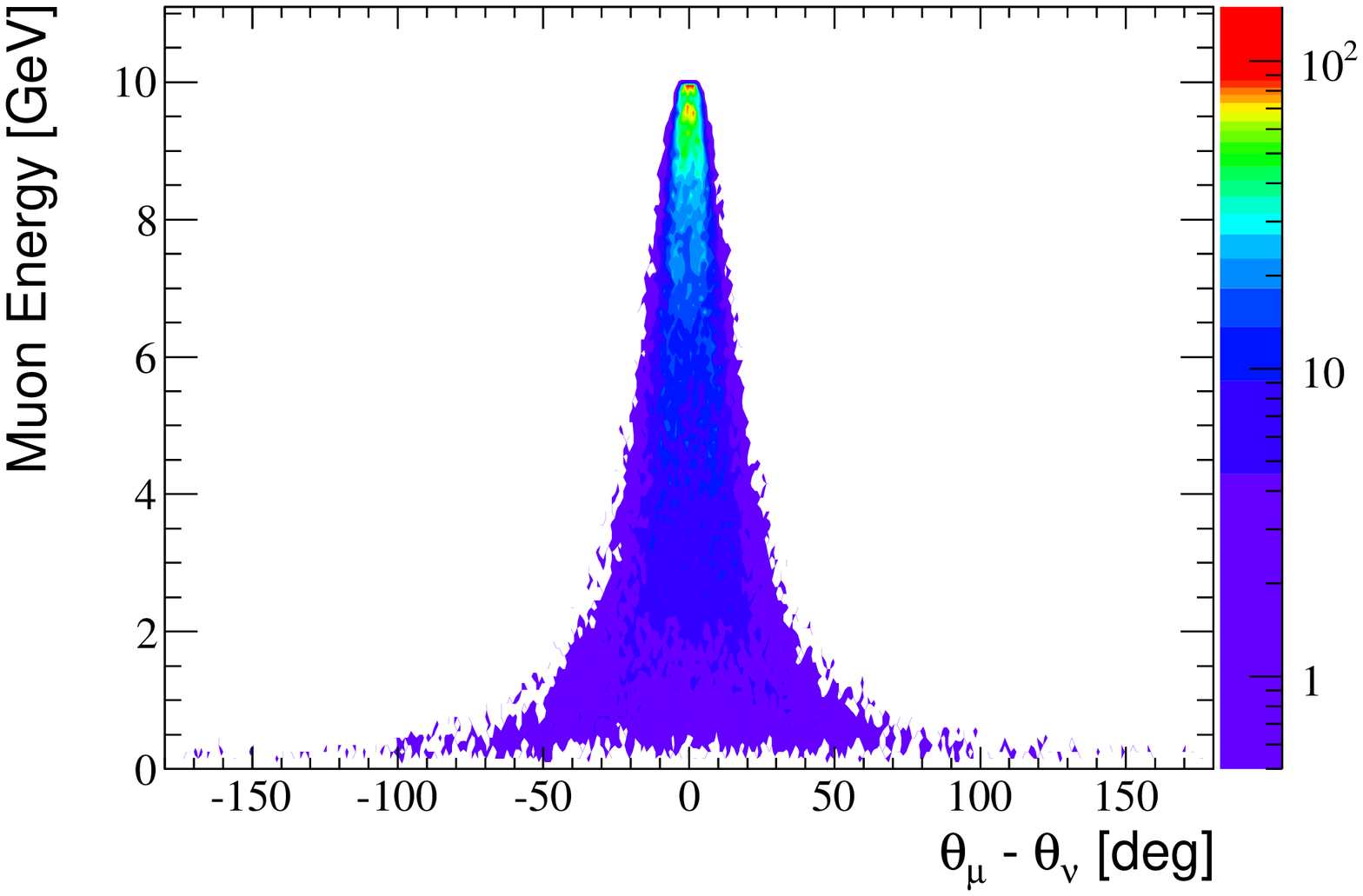}
\end{minipage}
\caption{Scatter plot of the muon energies and differences between the muon zenith angle ($\theta_{\mu}$) and the neutrino one ($\theta_{\nu}$), obtained for 10 GeV neutrinos.}
\label{fig:muangleene}
\hspace{0.5cm}
\begin{minipage}{0.6\linewidth}
\includegraphics[width=\columnwidth]{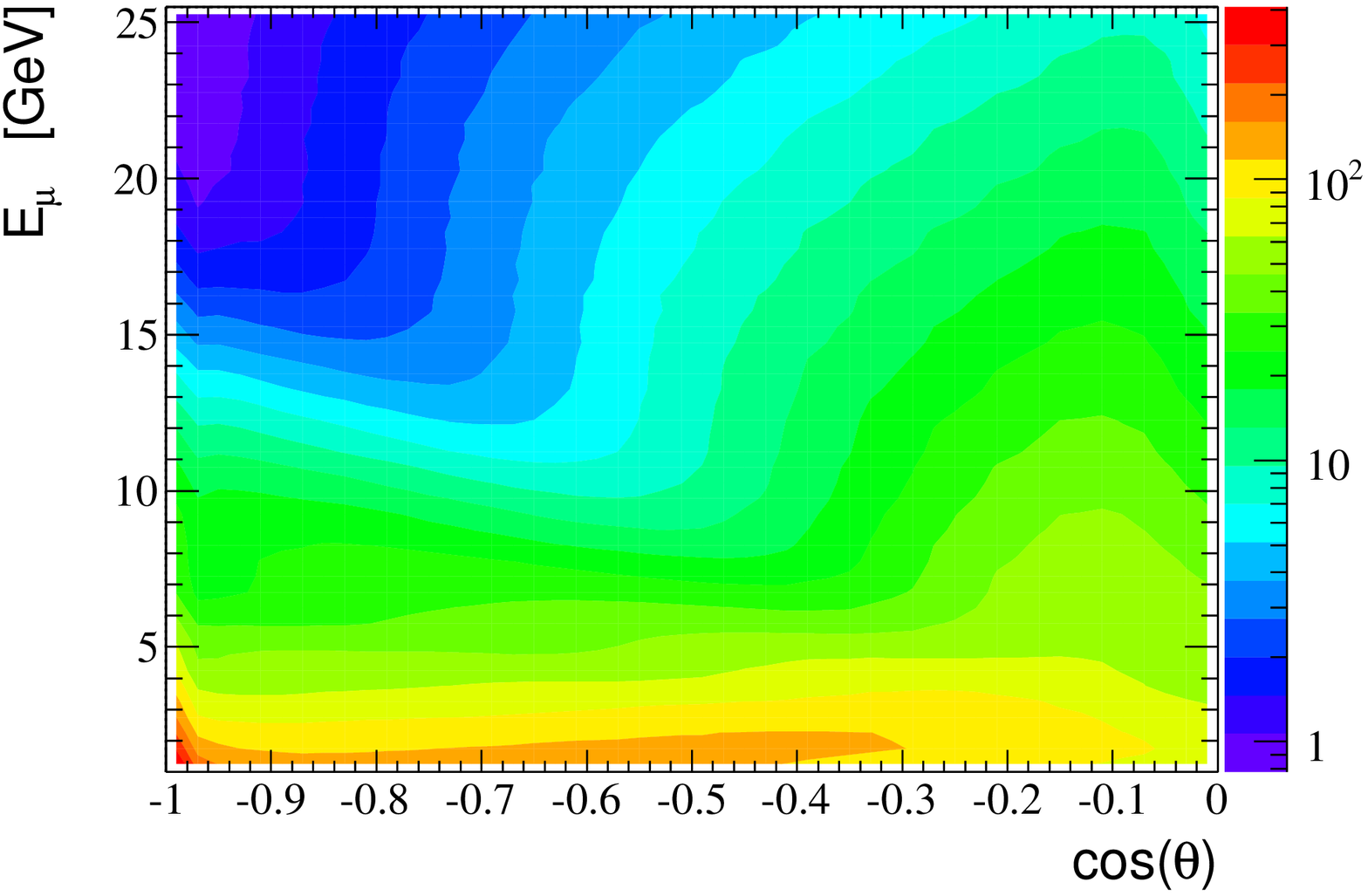}
\end{minipage}
\caption{Expected number of $\nu_{\mu}$ and $\bar{\nu}_{\mu}$ induced muon events in case of NH as a function of the energy and the cosine of the zenith angle.}
\label{fig:MatrixNHMu}
\end{center}
\end{figure}

To include the correct neutrino--muon kinematics, $\nu_{\mu}$ interactions were generated in an energy range from 1~GeV to 40~GeV (40000 events for each chosen energy in steps of 0.5 GeV) on a water target using the GENIE~\cite{Andreopoulos:2009rq} simulation code.
The zenith angle between the incoming neutrino and the outgoing muon in charged current interactions as well as its energy were computed.\\
An example of the results for 10 GeV neutrinos is shown in figure~\ref{fig:muangleene}.
According to the obtained distributions the muon energy and angle were randomly extracted for each neutrino event.

Although several studies are ongoing aiming at the exploitation of some coarse hadronic energy reconstruction in the neutrino energy determination, the conservative assumption that only the muon energy can be reconstructed is made in this work. Hence, the analysis was done using muon energy (E$_{\mu}$) instead of the neutrino one (E$_{\nu}$).\\
A threshold at 5~GeV was set in order to guarantee a reasonable energy and direction reconstruction. In addition, at lower energy the uncertainty on the cross sections increases up to 20\% and this additional systematics, neglected in the present work, should be considered. The energy range used in this work is therefore 5 to 40 GeV.\\
The matrices used in the MC Toy, to evaluate the NMH sensitivity, were obtained from those generated by GLoBES, applying the kinematical smearing and using the selected effective exposure. An example obtained using the central values for the oscillation parameters is shown in figure~\ref{fig:MatrixNHMu}.\\
Note that although those matrices are displayed in terms of $\cos \theta$ for a direct comparison with previous works,  the MC works directly in $\theta$ in order to treat correctly the angular smearing.

\begin{figure}[t]
\begin{center}
\includegraphics[width=0.6\columnwidth]{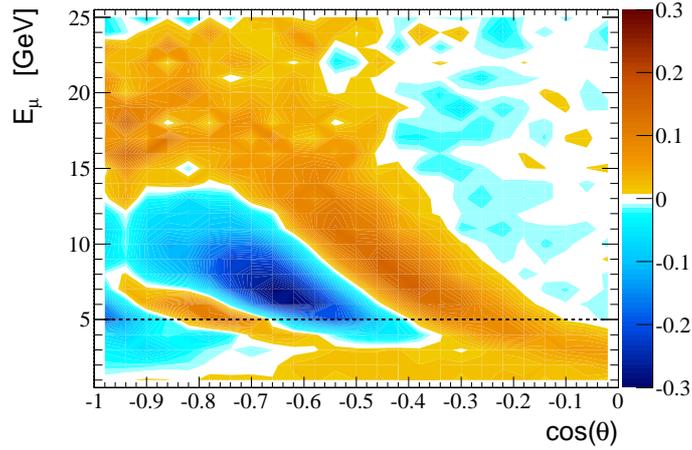}
\caption{Asymmetry (as defined by eq.~\ref{eq:asymmetry}) between the number of $\nu_{\mu}$ and $\bar{\nu}_{\mu}$ induced muon events expected in case of NH and IH, expressed as a function of the energy and the cosine of the zenith angle.}%, for an effective exposure of 34 Mt $\times $year.}
\label{fig:asymmetry}
\end{center}
\end{figure}

Based on the difference between the matrix generated in case of NH and IH, it is possible to identify the region where the effect is larger and therefore the discrimination more powerful.\\ 
The asymmetry defined as: 
\begin{equation}
\label{eq:asymmetry}
2 \times \frac{M_\mathrm{NH}-M_\mathrm{IH}}{M_\mathrm{NH}+M_\mathrm{IH}},
\end{equation}
shown in figure~\ref{fig:asymmetry}, was chosen as figure of merit, where $M_\mathrm{NH}$ and $M_\mathrm{IH}$ are the number of expected events at a given angle and energy for NH and IH respectively.\\
The region where the effect is more evident is between 5 and 10~GeV. Hence, the development of a detector with high  energy resolution at low energies is mandatory.
%\begin{figure}[htbp]
%\begin{center}
%\includegraphics[width=0.5\columnwidth]{./figure3bis.eps}
%\caption{Expected number of $\nu_{\mu}$ and $\bar{\nu}_{\mu}$ induced muon events in case of NH as a function of the energy and the cosine of the zenith angle.}
%\label{fig:MatrixNHMu}
%\end{center}
%\end{figure}
\label{MC}

\section{Reference oscillation parameters}
%A very important point in the construction of the matrixes that will be used in the following to evaluate the discrimination sensitivity between the two hierarchies, is the choice of the oscillation parameters.\\

\begin{table}[b]
\centering
\begin{tabular}{ll}
\hline
Parameter & Value \\
\hline
$\Delta m^2_{21}$ \cite{Beringer:1900zz} & (7.58$^{+0.22}_{-0.26}$)$\times 10^{-5}$ eV$^2$   \\
$\Delta m^2_{31}$(NH) \cite{Schwetz:2011qt}&  (2.45 $\pm$ 0.09)$\times $10$^{-3}$ eV$^2$ \\
$\Delta m^2_{31}$(IH) &  0.13$\times$10$^{-3}$ eV$^2$ -  $\Delta m^2_{31}$(NH)\\
$\sin^2(2\theta_{12})$ \cite{Beringer:1900zz} & 0.849 $^{+0.071}_{-0.059}$ \\
$\sin^2(2\theta_{13})$ \cite{Machado:2012} & 0.096 $\pm$ 0.013\\
$\sin^2(2\theta_{23})$ \cite{Beringer:1900zz} & 0.974$^{+0.026}_{-0.032}$\\
\hline
\end{tabular}
\caption{Central values and 1$\,\sigma$  uncertainty of the oscillation parameters used in this work.}
\label{tab:osc}
\end{table}

\begin{figure}[ht]
\begin{center}
\includegraphics[width=0.6\columnwidth]{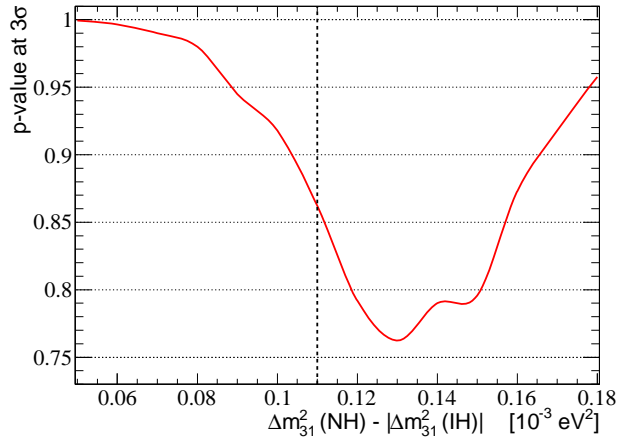}
\caption{Probability of discriminating between NH and IH at 3$\,\sigma$ C.L. as a function the difference in the values of $\Delta m^2_{31}$ assumed for NH and IH. The plot stands for an effective exposure of 34 Mt $\times$ year. The dashed vertical line corresponds to the best fit from ref.~\cite{Schwetz:2011qt}.}
\label{fig:scandm}
\end{center}
\end{figure}

The oscillation parameter central values, quoted in table~\ref{tab:osc}, and to  $\delta_{CP}$=0, are assumed in this work as reference values, except when specified.\\
The mixing angle $\theta_{23}$ has been extensively explored in the first octant, which is preferred at 1.5$\, \sigma$ by the global fit of ref.~\cite{GonzalezGarcia:2012sz}  in normal hierarchy. The results obtained in this work were cross checked by testing $\theta_{23}$  values in the second octant, without observing notable differences.\\
In the formalism adopted in this work,  the  best fit in normal hierarchy of the largest  $\Delta m^2$  is assigned to $\Delta m^2_{31}$(NH).  The value of $\Delta m^2_{31}$(IH) then differs from $\Delta m^2_{31}$(NH) by:
\begin{eqnarray}
\delta m^2_{31} & = & \Delta m^2_{31}(\mathrm{NH}) - |\Delta m^2_{31}(\mathrm{IH})|  \\
& =  & 2 \Delta m^2_{21} (\cos^2\theta_{12} - \cos\delta_{CP} \sin\theta_{13} \sin 2\theta_{12} \tan\theta_{23}) \nonumber
\end{eqnarray}
as pointed out in ref. \cite{Nunokawa:2005,Blennow:2012}. The dependence of $\delta m^2_{31}$ on $\delta_{CP}$ makes its value non--univocally assigned.  However, contrarily to the standard studies on the mass hierarchy discrimination with fit procedures,  where $\Delta m^2_{31}$(IH) is a free parameter, in the Monte Carlo Toy approach $\Delta m^2_{31}$(IH)  must be fixed. To overcome this problem, a dedicated study has been performed,  varying $\delta m^2_{31}$  from 0.05 to 0.18 $\times$ 10$^{-3}$ eV$^2$, and evaluating the p-value at 3~$\sigma$ with 34  Mt $\times$ year at 40 GeV of effective exposure (larger exposure would not allow to observe any effect). As it can be seen in figure~\ref{fig:scandm}, a minimum  was  found in 0.13$\times$10$^{-3}$ eV$^2$ (p-value $\sim$ 0.76), slightly  shifted with respect to the best value of $\sim$0.11$\times$10$^{-3}$ eV$^2$ from Ref.~\cite{Schwetz:2011qt}.

The value of $\delta m^2_{31}$ was conservatively fixed in this work to the so--obtained minimum. The effect induced by the $\Delta m^2_{31}$ shift between NH and IH becomes negligible (p-value $\sim$1) from an effective exposure of $\sim$100 Mt $\times$ year at 3$\,\sigma$.

\label{Introsys}

\section{Exposure}
\label{Exposure}
\label{sec:exposure}

\begin{figure}[t]
\begin{center}
\begin{minipage}{0.6\linewidth}
\includegraphics[width=\columnwidth]{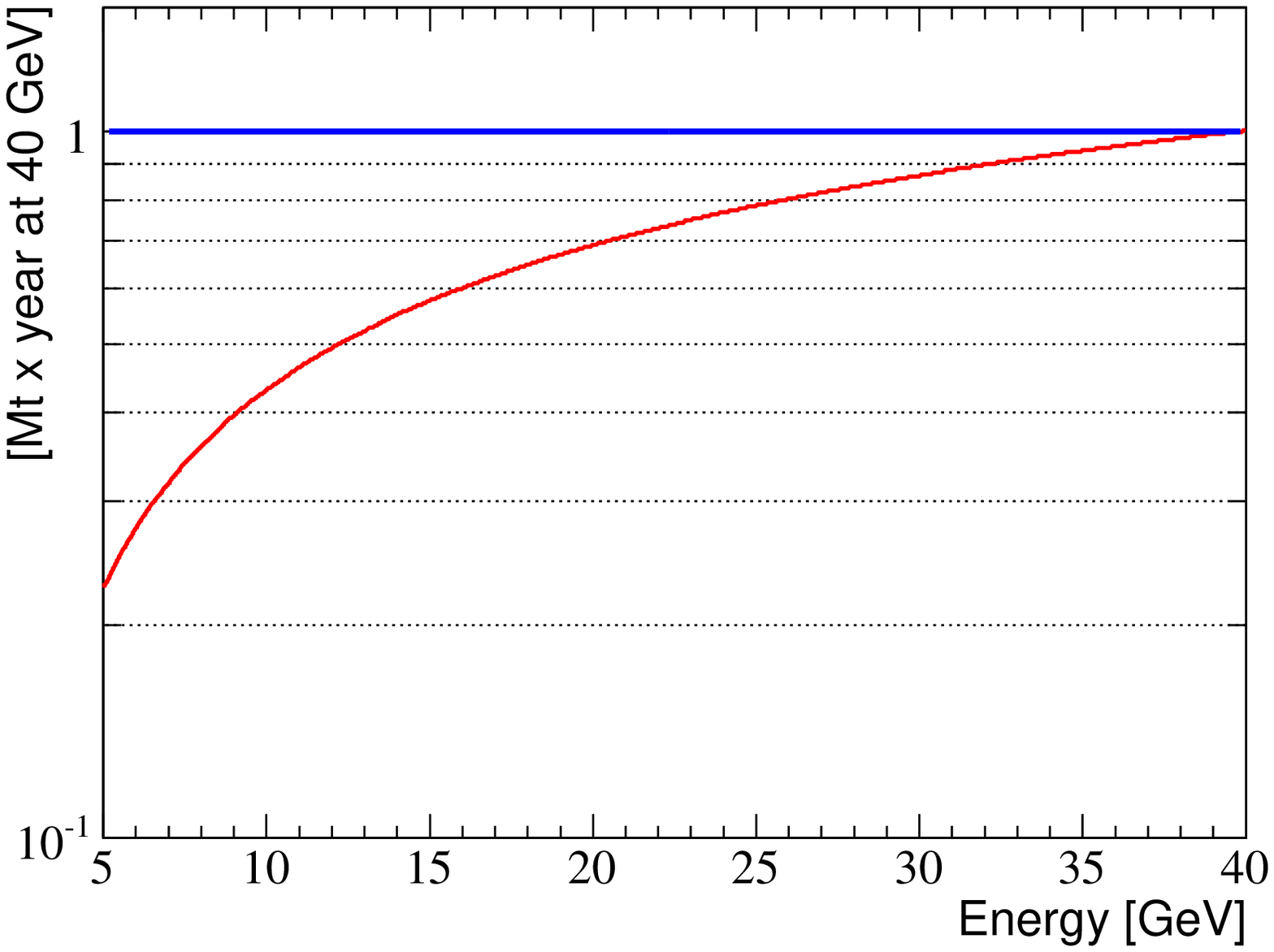}
\end{minipage}
\caption{Energy dependent (red) and optimal (blue) exposure profile, normalized at 1 Mt $\times$ year at 40 GeV.}
\label{fig:Mass}
\hspace{0.5cm}
\begin{minipage}{0.6\linewidth}
\includegraphics[width=\columnwidth]{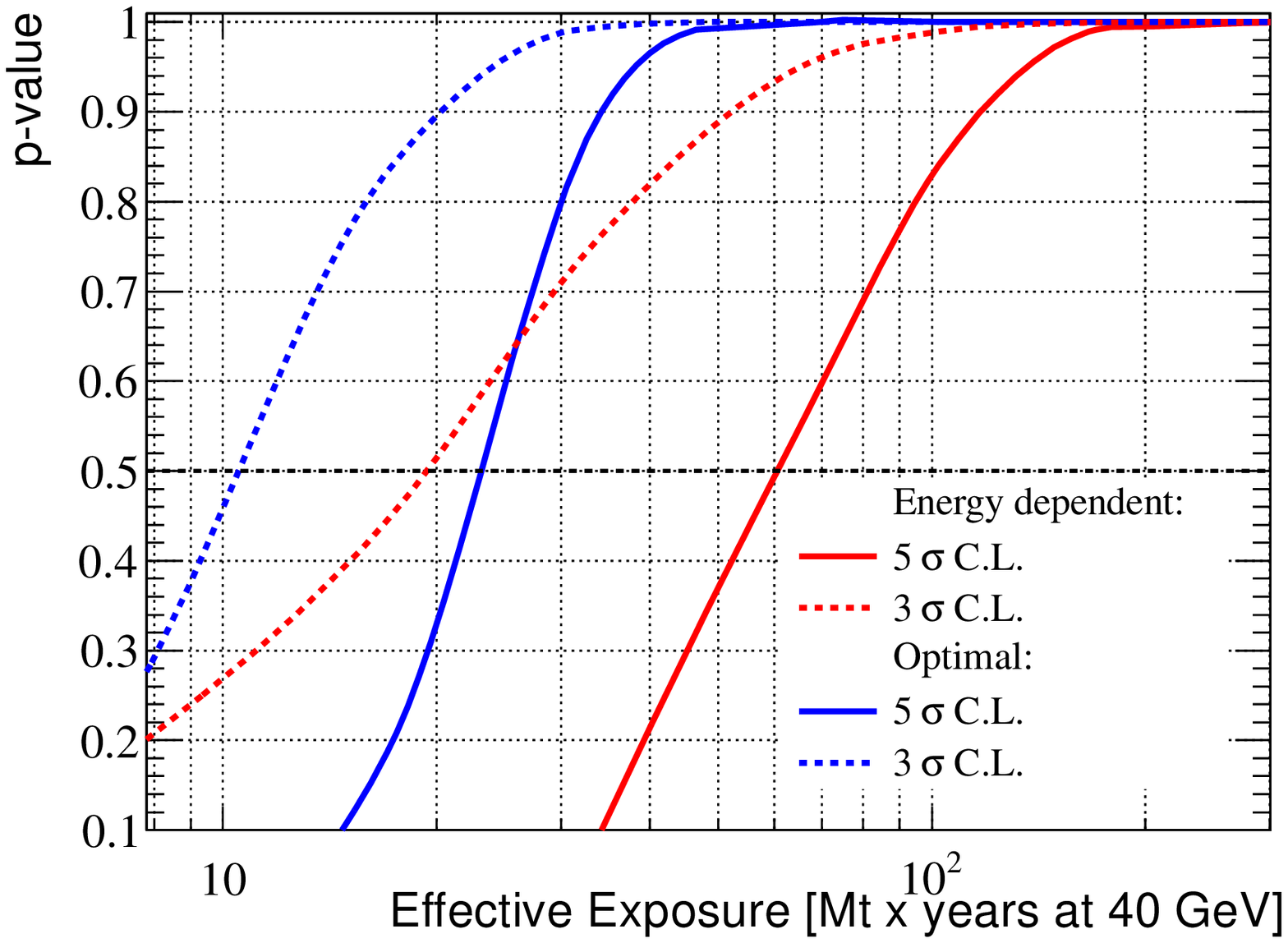}
\end{minipage}
\caption{p-value for 3$\,\sigma$ (dashed lines) and 5$\,\sigma$ (solid lines) C.L. as a function of the effective exposure for an ideal detector, assuming energy dependent (red) and optimal (blue) mass profiles. The horizontal dashed line is a guideline to read the required  exposures for a 50\% chance to discriminate between the two hierarchies at the corresponding C.L..}
\label{fig:exposure}
\end{center}
\end{figure}

%\begin{figure}[t]
%\begin{center}
%\includegraphics[width=0.6\columnwidth]{./figure6.eps}
%\caption{Energy dependent (red) and optimal (blue) exposure profile, normalized at 1 Mt $\times$ year at 40 GeV.}
%\label{fig:Mass}
%\end{center}
%\end{figure}
%
%\begin{figure}[h]
%\begin{center}
%\includegraphics[width=0.6\columnwidth]{./figure7.eps}
%\caption{p-value for 3$\,\sigma$ (blue) and 5$\,\sigma$ (red) C.L. as a function of the effective exposure for an ideal detector, assuming energy dependent (solid lines) and optimal (dashed lines) mass profiles. The horizontal dashed line is a guideline to read the required  exposures for a 50\% chance to discriminate between the two hierarchies at the corresponding C.L..}
%\label{fig:exposure}
%\end{center}
%\end{figure}

In the following, the impact of the model and parameter assumptions on the NMH discrimination power is evaluated relatively to a starting ideal condition.
For this ideal case, perfect energy and angle resolutions, the reference conditions, mentioned in the previous section, and the energy-dependent mass profile of eq.~\ref{eq:mass}, shown in figure~\ref{fig:Mass}, are assumed. At this stage, no biases in the true models parameters are introduced. \\
Setting the p--value threshold at 0.5 at 5$\,\sigma$ C.L., the minimal required effective exposure is 60 Mt $\times$ year, as shown in figure~\ref{fig:exposure}. \\
As already mentioned, there is no currently  available detailed studies on  ORCA/PINGU mass profiles.  Other references, e.g. \cite{DeYoung:2011},  quote different profiles with respect to eq.~\ref{eq:mass}~\cite{Akhmedov:2012ah}. To understand the impact of the mass profile, the optimal case with neutrino detection efficiency equal to 1, independently on the energy, was  tested. The corresponding effective exposure profile and p--value are shown in figure~\ref{fig:Mass}  and \ref{fig:exposure}, respectively.  In this case, the required effective exposure, for a p-value equal to 0.5 at 5$\, \sigma$ C.L., is reduced by a factor $\sim$3. This result demonstrates the significant impact of the detection efficiency in the lower energy region, namely at 5--10~GeV, where the NMH asymmetry is large, as shown in figure~\ref{fig:asymmetry}, and the expected flux is high. Future Monte Carlo studies should focus on identifying the detector configuration able to optimize the efficiency at lower energies.\\

In order to better appreciate the impact of the model uncertainties and of the detector resolutions, discussed in the next sections, an effective exposure of 170 Mt  $\times$ year, corresponding to a p--value $\sim$1 at 5$\,\sigma$ C.L.  is hereafter assumed.

\section{Flux models}
\label{sec:flux}
\label{sec:fluxsys}

\begin{figure}[t]
\begin{center}
 \begin{minipage}{0.6\linewidth}
\includegraphics[width=\columnwidth]{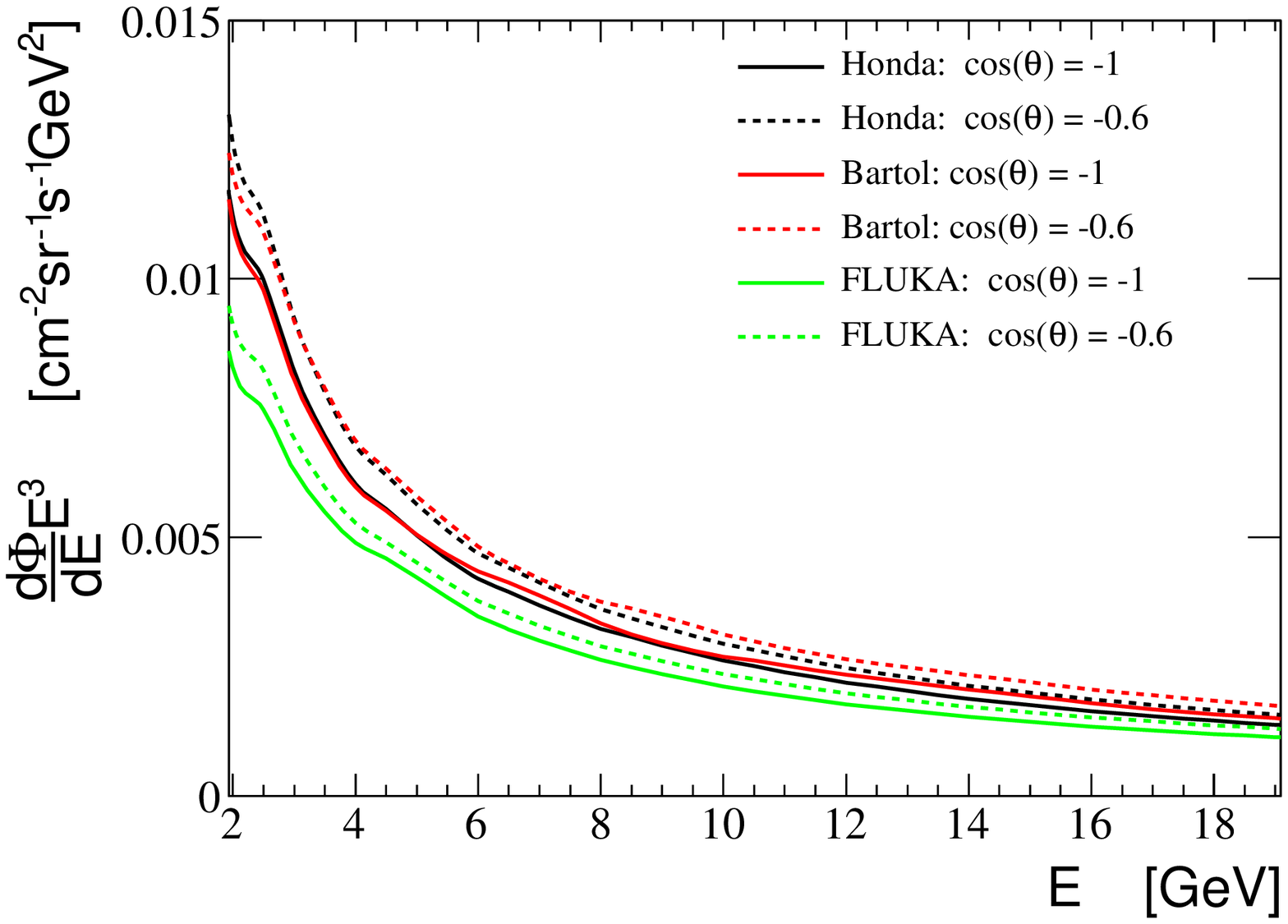}
\end{minipage}
\caption{Energy dependance of the atmospheric $\nu_\mu$ differential flux for 3 different models: Honda (black), FLUKA (green) and Bartol (red). Solid lines represent the vertical component of the flux whereas dashed lines correspond to the fluxes incident at $\cos\theta=-0.6$.}
\label{fig:fluxes_abs}
\hspace{0.5cm}
\begin{minipage}{0.6\linewidth}
\includegraphics[width=\columnwidth]{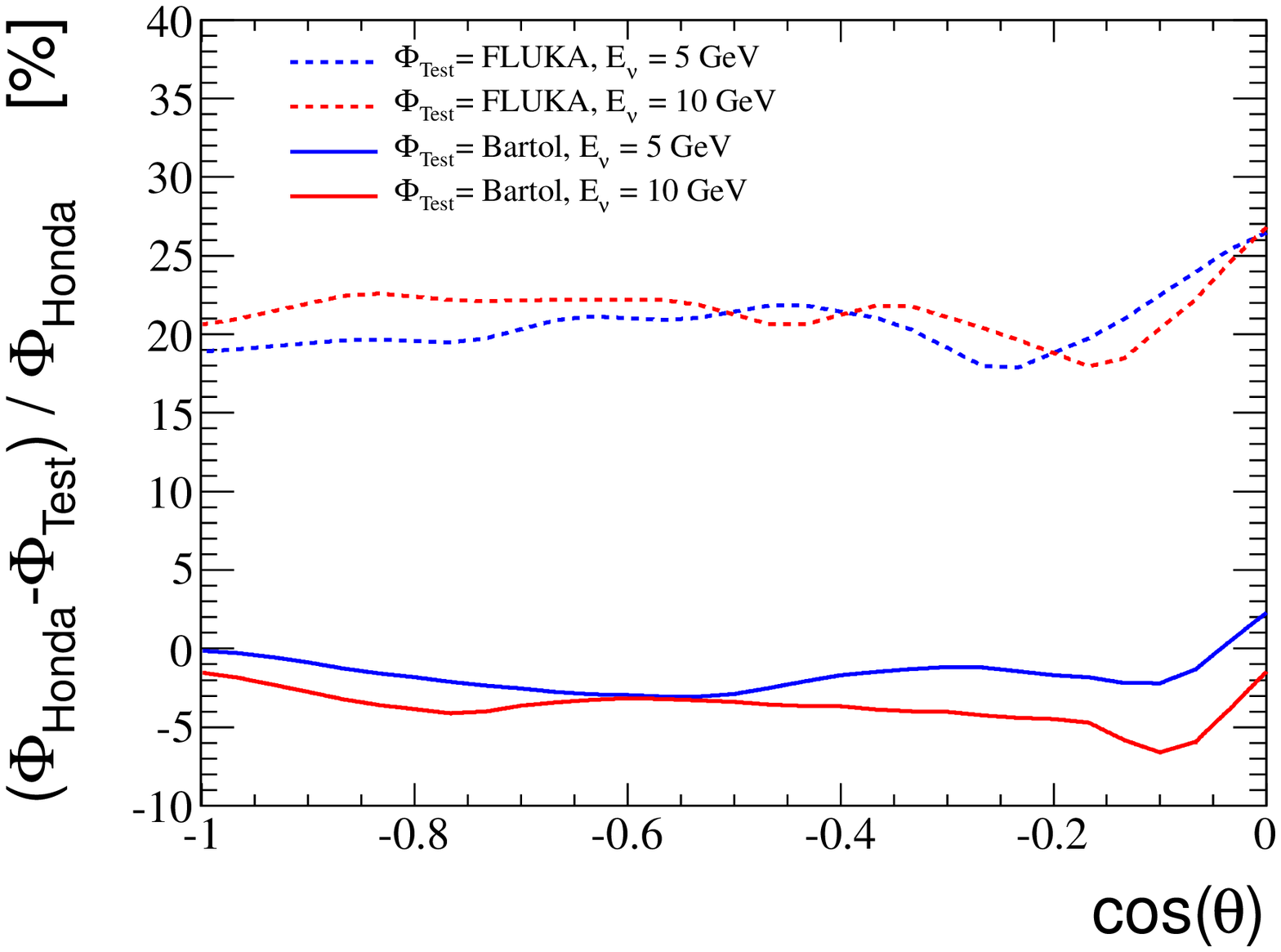}
\end{minipage}
\caption{Difference between atmospheric $\nu_\mu$ fluxes as predicted by Honda and Bartol (solid lines), and by Honda and FLUKA (dashed lines). Blue lines correspond to 5 GeV neutrinos whereas red lines represent 10 GeV neutrinos.}
\label{fig:fluxcomp}
\end{center}
\end{figure}

\begin{figure}[t]
\begin{center}
\begin{minipage}{0.6\linewidth}
\includegraphics[width=\columnwidth]{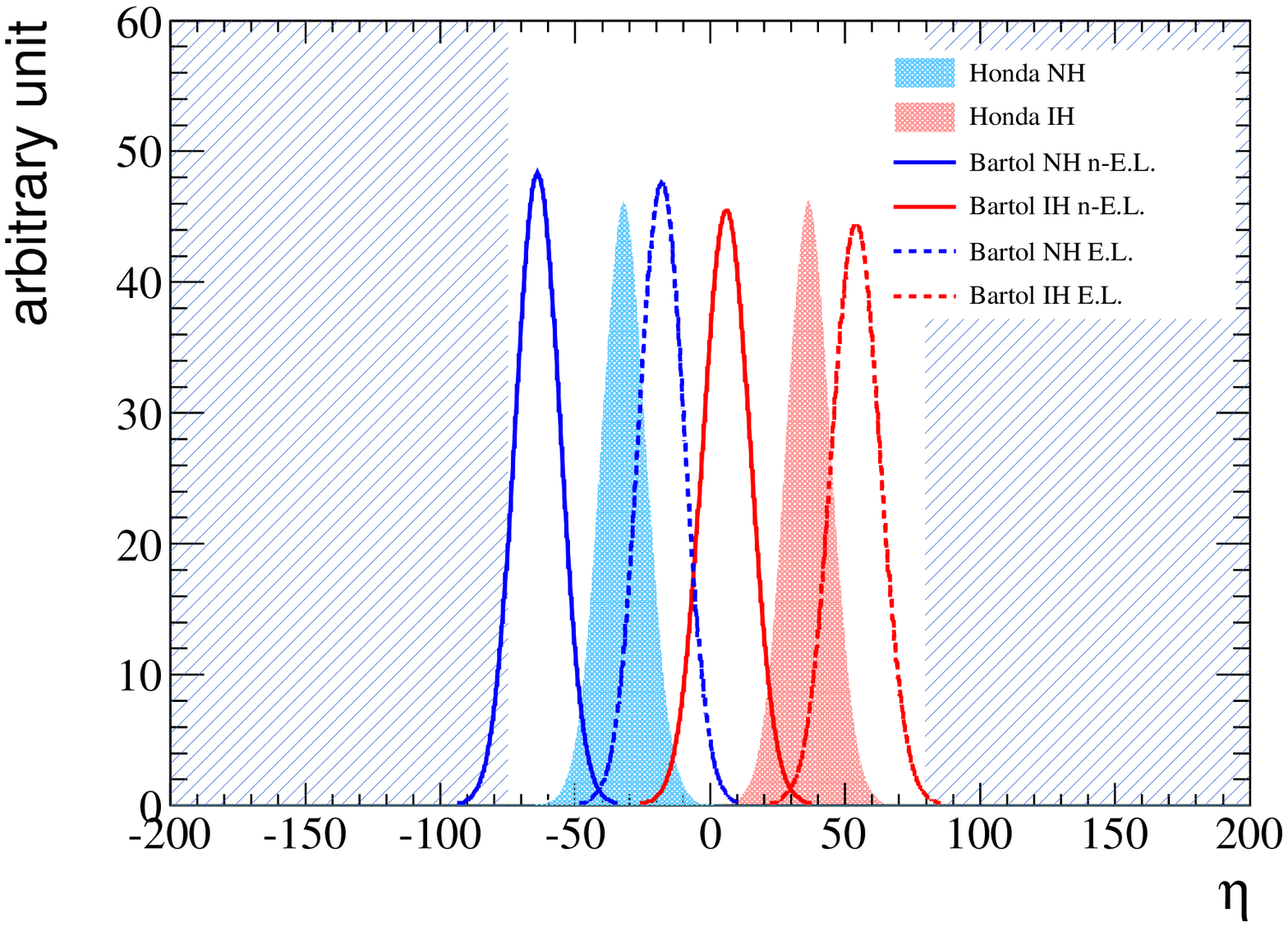}
\end{minipage}
\hspace{0.5cm}
\begin{minipage}{0.6\linewidth}
\includegraphics[width=\columnwidth]{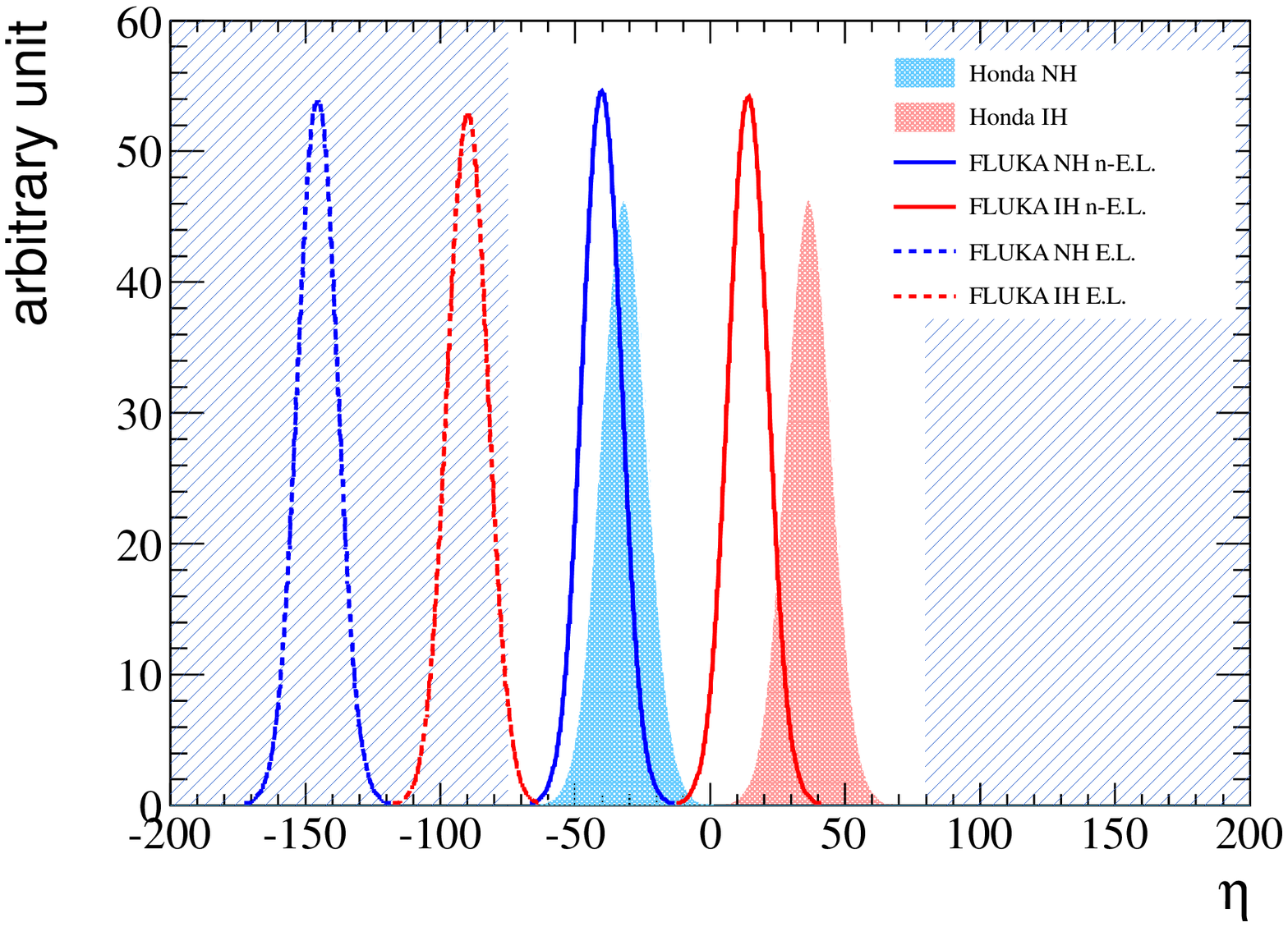}
\end{minipage}
\caption{$\eta$ distributions (top) assuming Honda as model hypothesis and Honda (shaded area) or Bartol (lines) as true hypothesis. Solid lines corresponds to the non-extended likelihood (n-E.L.), and dashed lines to the extended one (E.L.). The same results are shown (bottom) for  the FLUKA flux, instead of the Bartol one. }
\label{fig:sysFlux}
 \end{center}
\end{figure}

\begin{figure}[h]
\begin{center}
\includegraphics[width=0.6\columnwidth]{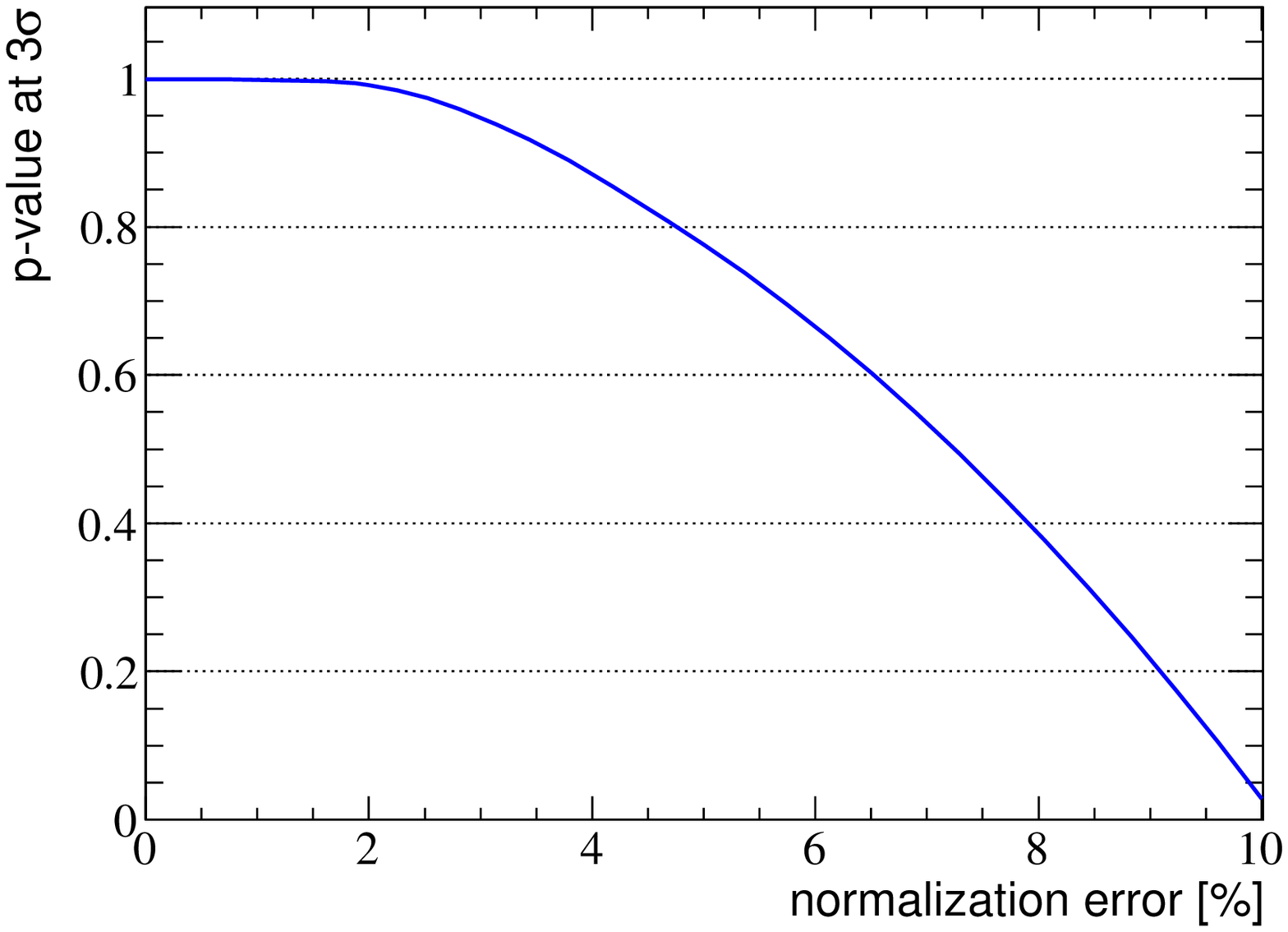}
\caption{p-value computed at 3$\,\sigma$ with the extended unbinned likelihood as a function of the Honda flux normalization error.}
\label{fig:testlikelihood}
\end{center}
\end{figure}

Available predictions on the atmospheric neutrino fluxes show differences both in the absolute normalization and in the energy and angular profiles.\\ 
 In this section, both the maximum acceptable uncertainty on the flux amplitude and the impact of the flux profile are evaluated. The reference model used in this work, Honda 1995~\cite{Honda:1995}, was compared with other predictions: FLUKA 2002~\cite{Battistoni:2002ew} and Bartol 1995~\cite{Agrawal:1995gk}.  A detailed description of the three models and of their differences  can be found in ref.~\cite{Stanev:2004}.  Recent model updates~\cite{Honda:2011nf} did not significantly modify the respective flux profile and amplitude predictions.  \\
Energy (figure~\ref{fig:fluxes_abs}) and angular (figure~\ref{fig:fluxcomp})  dependences of the three fluxes agree in  shape,  above 5~GeV, at the $\sim$5\% level. In terms of amplitude, whereas the Honda and Bartol models are in agreement at the level of a few percents, the FLUKA model differs from the others by more than 20\%.

To quantify the impact of the flux uncertainties on the p-value, a bias in the true hypothesis  was applied, as described in  section~\ref{sec:method}: the Honda flux was assumed as the model hypothesis, whereas the FLUKA and the Bartol ones were used as biased true hypotheses.\\
The slight discrepancy between Honda and Bartol fluxes lead to a relatively small loss in the discrimination power, at the level of p-value of 0.851, at 5$\,\sigma$ C.L.. On the contrary,  the disagreement between Honda and FLUKA was found more significant, lowering the correspondent p-value down to 0. The two cases are shown in figure~\ref{fig:sysFlux} (dashed lines), where the biased test statistics are compared with the unbiased ones.

To demonstrate that the impact is mostly due to the differences in normalization, the extended unbinned likelihood of eq.~\ref{eq:likelihood}  was  substituted with the following non-extended one:

 \begin{equation}
 \label{eq:like_nonextended}
L_j =  \prod_{i=1}^n \mathrm{pdf}_j(E_i,\theta_i)
\end{equation}
where $j=\{$NH, IH$\}$, and $i=1...n$ is the event index.\\
Removing the extended component of the unbinned likelihood from eq. \ref{eq:likelihood}, the test statistic becomes  independent on the expected number of events. In general, this approach has the consequence to weaken the discrimination power, by limiting  the  constraints to the energy and angular shapes only. In the unbiased case, where the Honda flux is assumed in both the model and true hypotheses,  p-value reduces to  0.985.\\
When testing Bartol versus Honda in the non-extended case, the discrimination power lowers down to a p-value equals to  0.502, while when  testing FLUKA versus Honda,  p-value rises up from 0 in the extended case to 0.655. These results, shown in figure~\ref{fig:sysFlux}, demonstrate that the FLUKA profile is relatively more similar to the Honda profile, than to the Bartol one.

The natural question arising from this study regards the convenience of using the extended with respect to the non-extended unbinned likelihood. The answer depends on the uncertainty that can be reached in the normalization factor.  The impact of the flux amplitude uncertainty on the discrimination power can be attenuated by anchoring the fit to a region in the  parameter space where the  effect of mass hierarchy is minimized. An example, as shown in figure~\ref{fig:asymmetry}, is provided by the region corresponding to almost horizontal ($\cos \theta \sim 0$) high-energy (E$_{\nu} > 20$ GeV) neutrinos.  \\
Assuming the Honda flux in both the  model and true hypotheses, the p--value of the extended unbinned likelihood was computed  by biasing the normalization factor in the true case from 0 to 10\%, as shown in figure~\ref{fig:testlikelihood}.\\  The p-value is not sensitive to variations in the  normalization up to 2\%.  Above this value, the p-value rapidly decreases to 50\% for a bias of  7.5\%, and reaches almost 0 with a normalization error of 10\%. Assuming a minimum  threshold of the p-value at 50\%, the extended unbinned likelihood estimator is considered convenient, if the uncertainty on the flux normalization is below 7.5\%. Above this value, the non-extended estimator offers a more robust discrimination power.

\section{Earth density profile}
\label{sec:profsys}

\begin{figure}[t]
\begin{center}
\includegraphics[width=0.6\columnwidth]{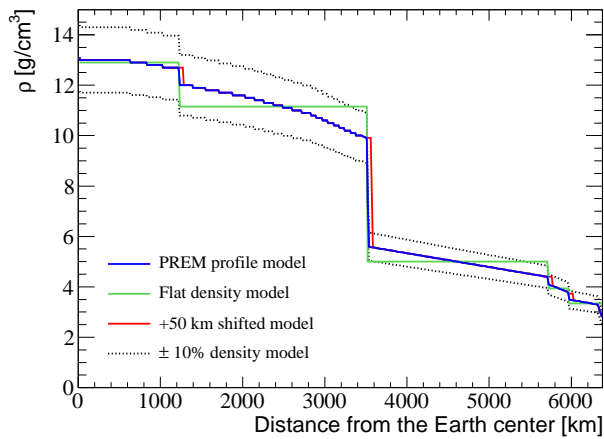}
\caption{Earth matter density profile as a function of the distance from the  Earth center for the PREM model (blue), the modified flat density model (green), the model with shifted zone boundaries (red) and the model with densities modulated by $\pm 10$\% (dashed black).}
\label{fig:Stacey}
\end{center}
\end{figure}

Uncertainties on the Earth density profile can play a significant role in neutrino matter oscillations. Recent studies (see ref.~\cite{Agarwalla:2012uj} and references therein) indicate that density values predicted by the PREM, averaged over  100 km, can be assumed to be known at the level of a few per cent at all depths in the core and mantle (the crust can be neglected, as it represents a very small fraction of matter traversed by neutrinos), while global variations of the density are constrained at the 10$^{-4}$ level.  The Earth ellipticity also has an impact on the matter density profile along the neutrino propagation.

In order to understand the effects of the PREM--related parameters  on the p-value, the following modifications were independently introduced: the density was varied by an overall factor of $\pm$10\%,  the boundaries of each zone shifted by  50 km, and a flat density profile between the main discontinuities was assumed, as shown in figure~\ref{fig:Stacey}. The introduced biases are larger than the known uncertainties or even  unphysical, but needed to appreciate the PREM parameters impact, otherwise below the sensitivity of the  statistical method adopted here.

Both the 50 km limit shift and the assumption of the flat density profile have almost no impact on the mass hierarchy discrimination power, with a p-value at 3$\,\sigma$ equals to 0.999. The p-value is almost unchanged (0.996) when reducing the overall density  by 10\%,  and it is slightly affected (0.872) when density is increased by same factor. No impact was observed by varying the overall density factor by 1\%, in agreement with the error found in literature~\cite{Agarwalla:2012uj,Bolt:1991}.

In conclusion, uncertainties on the PREM parameters have a negligible impact on the mass hierarchy determination.

\section{Neutrino oscillation parameters}
\label{sec:oscilsys}

\begin{figure}[]
 \begin{center}
\includegraphics[width=\textwidth]{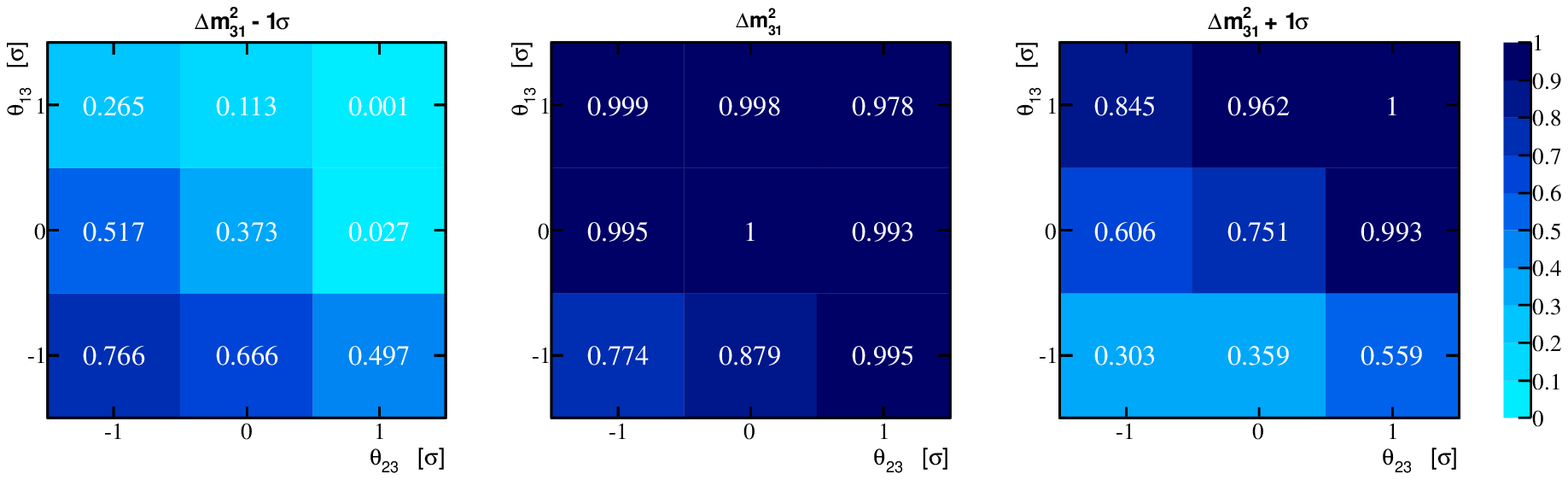}
\includegraphics[width=\textwidth]{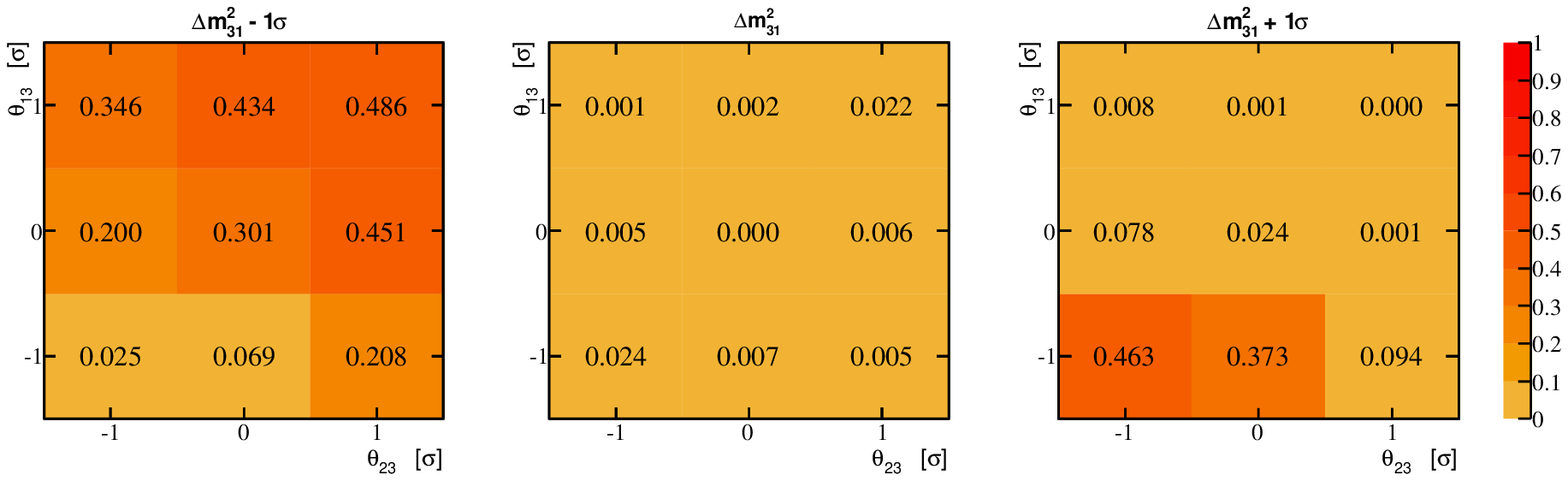}

\caption{p-value (top) and fraction of false positives (bottom) at 3$\,\sigma$ C.L. as a function of the $\theta_{13}$ and $\theta_{23}$ parameter variations at $\pm 1\, \sigma$ for $\Delta m^2_{31}$ shifted by $-1\, \sigma$ (left), no shift (middle) and by $+1\, \sigma$ (right).}
\label{fig:AtmSys}
 \end{center}
\end{figure}

\begin{figure}[]
\begin{center}
\includegraphics[width=\textwidth]{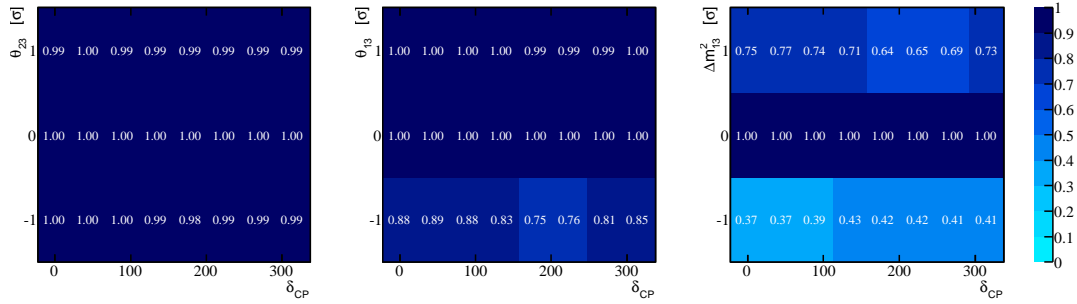}
    
\caption{p-value at 3$\, \sigma$ C.L. as a function of $\delta_{CP}$ and $\theta_{23}$ (left), $\theta_{13}$ (middle) and $\Delta m^2_{31}$ (right) parameter variations.}
\label{fig:CPSys}
\end{center}
\end{figure}

A careful study aimed to identify critical oscillation parameters, and their correlation, is a necessary condition for future analyses.\\  
The sensitivity study follows  the scheme adopted in the previous sections: each oscillation parameter value is biased in the true hypothesis, by  $\pm 1\, \sigma$ from the central value, as quoted in table~\ref{tab:osc}, while keeping unaltered the model hypothesis.

As already explained in section~\ref{Introsys}, the shift between the absolute values of $\Delta m^2_{31}$(NH) and $\Delta m^2_{31}$(IH), labelled as $\delta m^2_{31}$ has a negligible impact with an  effective exposure of 170 Mt$\times$ year. However, the p-value can decrease, biasing simultaneously  $\delta m^2_{31}$  and other oscillation parameters. The expected main correlation between the shift and $\Delta m^2_{31}$(NH) was therefore tested: with an effective exposure of 34 Mt $\times$ year, the p-value has a maximum variation of $\pm$8\%, while with an exposure of 170 Mt $\times$ year, the p-value spread is at the per-mill level. As a consequence, in the following analysis $\delta m^2_{31}$ will be fixed. From now on, $\Delta m^2_{31}$(NH) will be referred as $\Delta m^2_{31}$.

Other parameters, with an expected low impact, are the solar ones. The correlation study between $\theta_{12}$ and $\Delta m^2_{12}$ shows an almost constant p-value, with a maximum spread of 0.1\%, around 1, at 3$\, \sigma$ C.L.. Moreover, no degradation in the hierarchy discrimination was observed by studying each of the solar parameters in combination with $\Delta m^2_{31}$. As for the $\delta m^2_{31}$ case, solar parameters can be fixed to their central values, assuming a negligible contribution of their uncertainties to the discrimination power.

The p-value  dependence on parameters, in the atmospheric and reactor sectors, has been studied by varying $\theta_{13}$, $\theta_{23}$ and $\Delta m^2_{31}$ in all their possible combinations. The p-values at 3$\, \sigma$, as well as the fraction of false positives, are shown in figure~\ref{fig:AtmSys}. The strongest impact, as  foreseen, is due to $\Delta m^2_{31}$. Clear correlations are observed between  $\theta_{13}$ and $\theta_{23}$, for instance by shifting both the central values by -1$\, \sigma$. The effects induced biasing $\Delta m^2_{31}$ are, in some cases, compensated by shifts in $\theta_{13}$ and $\theta_{23}$. This is explained by the relative shifts of the biased $\eta$ distributions with respect to the unbiased one, as shown in figure~\ref{fig:pval}. In particular, biases on $\Delta m^2_{31}$ shift the distribution in the opposite direction with respect to $\theta_{13}$ and $\theta_{23}$.\\
When $\Delta m^2_{31}$ is fixed to its central value, the impact of  $\theta_{13}$ and $\theta_{23}$ biases are relatively modest (the minimum obtained p-value is equal to 0.774). However, the fraction of false positives for $\Delta m^2_{31} \pm 1\, \sigma$ strongly depends on $\theta_{13}$ and $\theta_{23}$ values, varying from a few per-mill to almost 50\%. This clearly demonstrates the non-negligible dependence of the mass hierarchy discrimination power on $\Delta m^2_{31}$, $\theta_{13}$ and $\theta_{23}$.

Finally, the dependence on $\delta_{CP}$ was studied by varying its value in steps of 45 degrees, in combination  with $\theta_{13}$, $\theta_{23}$ and  $\Delta m^2_{31}$. The results for the p-value at 3$\, \sigma$ C.L., are shown in figure~\ref{fig:CPSys}. The dependence on $\delta_{CP}$ is relatively weak and can be appreciated only for $\Delta m^2_{31} \pm 1\, \sigma$ and for $\theta_{13} - 1\, \sigma$, where the p-value varied by $\sim$16\% at the most.

In conclusion, a strong dependence of the NMH  discrimination power on $\theta_{13}$, $\theta_{23}$ and $\Delta m^2_{31}$ was observed, whereas a weak dependence was seen on $\delta_{CP}$. The effects induced by $\theta_{12}$, $\Delta m^2_{12}$, and by $\delta m^2_{31}$ are negligible.\\

\section{Detector resolution and energy threshold}
\label{sec:resolution}

\begin{figure*}[t]
\begin{center}
\includegraphics[width=0.7\textwidth]{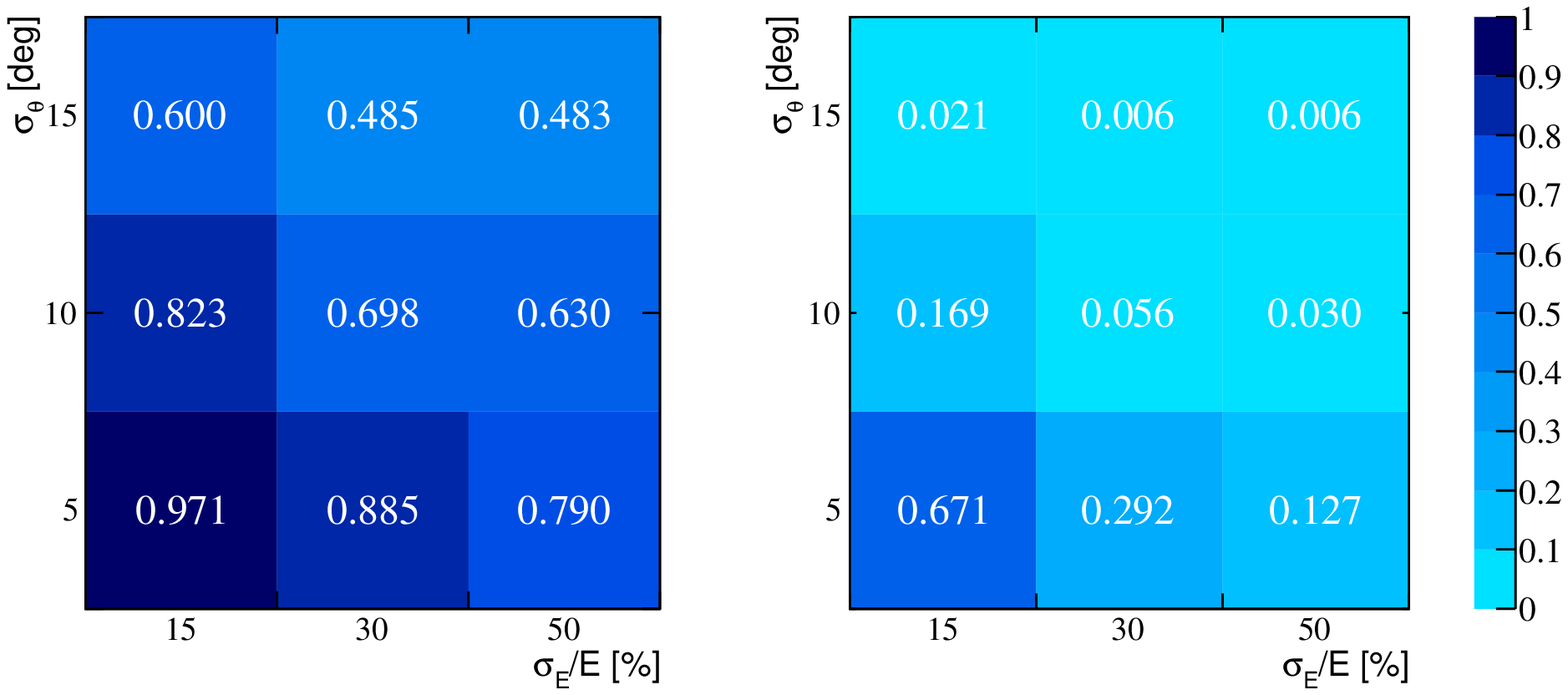}
\includegraphics[width=0.7\textwidth]{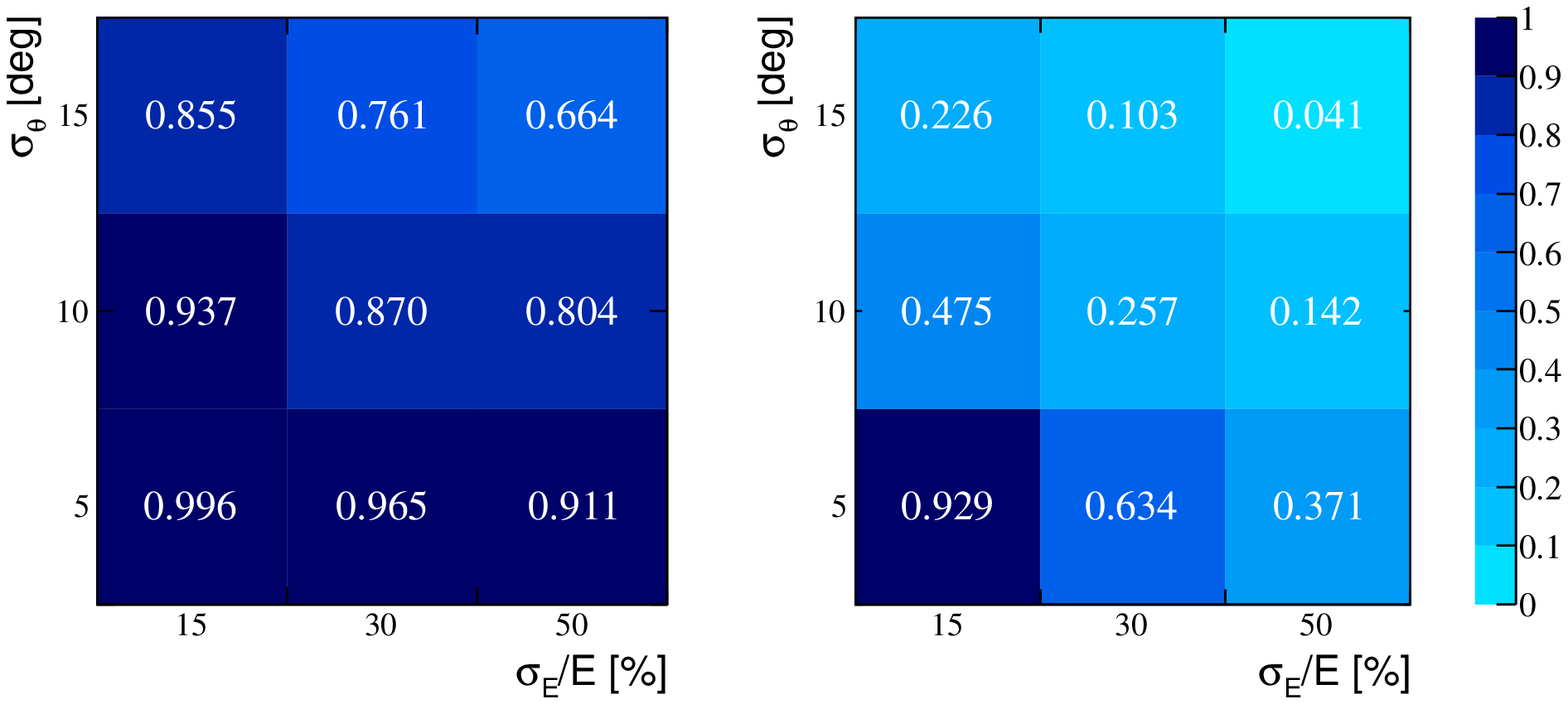}

\caption{p-value at 3$\,\sigma$ (left) and 5$\,\sigma$ (right) C.L. for different  detector energy and angular resolutions with 5~GeV (top) and 1~GeV (bottom)  muon energy  thresholds.  }
\label{fig:detres}
\end{center}
\end{figure*}

\begin{figure}[]
\begin{center}
\includegraphics[width=0.6\columnwidth]{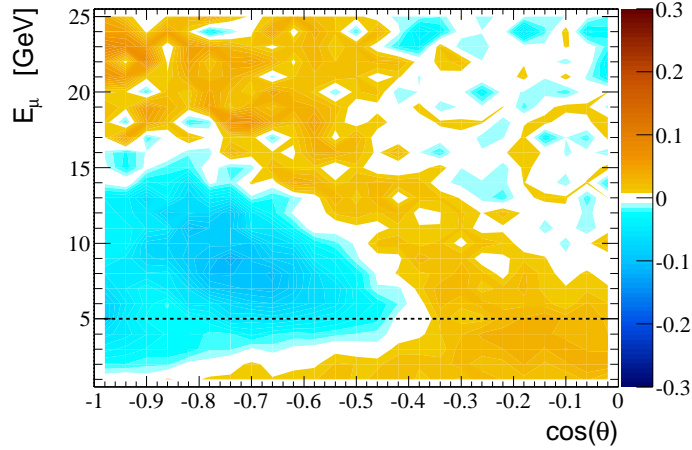}
\caption{Asymmetry (as defined by eq.~\ref{eq:asymmetry}) between the number of $\nu_{\mu}$ and $\bar{\nu}_{\mu}$ induced muon events expected in case of NH and IH, expressed as a function of the energy and the cosine of the zenith angle. A detector resolution of 10 degrees on the reconstructed zenith and an energy resolution of 30\% are assumed.}
\label{fig:asymmetry2}
\end{center}
\end{figure}

Finite detector resolutions introduce a further weakening of the discrimination power. To study their effects, both the true and the model hypotheses were modified the same way, according to the  angular and energy resolutions. A smearing was introduced on the energy and the zenith angle for each generated event in the Monte Carlo simulations, using Gaussian distributions. Events falling outside the analysis range, with $\cos \theta > 0$ or muon energy  $< 5$ or $> 40$~GeV, were discarded.

Since no Monte Carlo based prediction has been published so far with dedicated simulations on ORCA or PINGU detectors in the 5 to 40~GeV energy range, the following resolutions were arbitrarily tested: $\sigma_{\theta}$ = 5, 10 and 15 degrees, and $\sigma_{E}/E$  = 15\%, 30\% and 50\%.  
The so-obtained p-values are shown in figure~\ref{fig:detres} for a C.L. of 3 and 5$\, \sigma$. 
These values must be compared with the ones already evaluated in ideal conditions (see section~\ref{Exposure}): 1 and 0.992 for 3 and 5$\, \sigma$ C.L., respectively.

In figure~\ref{fig:asymmetry2} the NH--IH asymmetry is shown for $\sigma_{\theta}$ = 10 degrees and $\sigma_{E}/E$  = 30\%. Even if  the asymmetry has a pattern similar to the ideal case (see Fig.~\ref{fig:asymmetry}), the p-value lowers down to 0.698 at 3$\, \sigma$ C.L., and to 0.056 at 5$\, \sigma$ C.L.. The discrimination power can be partially recovered increasing the effective exposure.

Another way to increase the p-value, when reduced by detector resolutions, would be the reconstruction of low energy muons below the 5~GeV threshold.
The p-values obtained when the analysis range is 1 to 40 GeV are shown in figure~\ref{fig:detres}. A maximum increase of $\sim$28\% and $\sim$34\% on the p-value can be reached at 3 and 5$\, \sigma$ C.L. respectively. Nonetheless, in the low energy region between 1 and 5 GeV, degradations of the resolution are expected. In addition, as mentioned in section~\ref{sec:MC}, systematics coming from the cross sections uncertainty should be considered below 5~GeV. 

\begin{figure}[]
\begin{center}
\includegraphics[width=0.6\columnwidth]{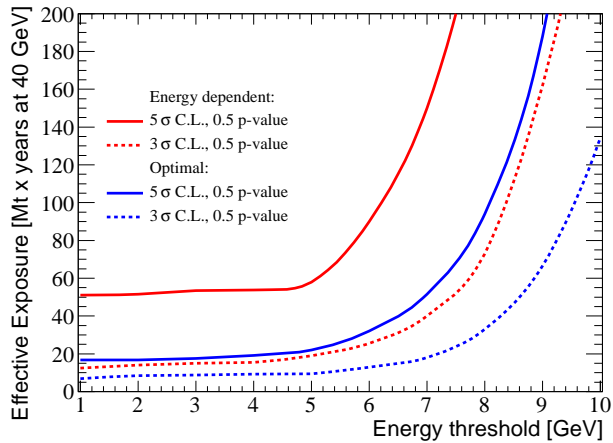}
\caption{
Exposure needed to reach a p-value of 0.5 at 3$\, \sigma$ (dashed lines) and 5$\, \sigma$ (solid lines) C.L.,  as a function of the muon energy threshold for an ideal detector, assuming energy dependent (red) and optimal (blue) mass profiles.}
\label{fig:threshold}
\end{center}
\end{figure}

For the sake of completeness, the exposure needed to reach a p-value of 0.5 at 3 and 5$\,\sigma$ C.L. is shown in figure~\ref{fig:threshold} as a function of the energy threshold, for an ideal detector. As expected,  thresholds larger than 5~GeV lead to a rapid reduction of the sensitivity.

%\input{Sensitivity}

% Put \label in argument of \section for cross-referencing

%\subsubsection{}

\section{Conclusions}
\label{sec:conclusion}

The oscillation pattern of atmospheric neutrinos in the 1--20~GeV energy range  can be exploited by future large volume ice/water Cherenkov detectors to discriminate between NH and IH. In this work, a Toy Monte Carlo, based on an unbinned likelihood ratio test statistic, was developed  to quantitatively investigate their discrimination potential.  This  approach allows to compute the probability of NMH discrimination,  at a given C.L., while keeping track of possible  misidentifications between hierarchies. For this purpose, a complete simulation chain, from neutrino generation to propagation and interaction, was implemented based on GLoBES and GENIE software tools. 

The discrimination power was evaluated as a function of the exposure in the ideal case of perfect detector resolutions and no systematics. Assuming  the same energy dependence of the detector effective mass as in ref.~\cite{Akhmedov:2012ah}, the minimum required exposure for a 50\% discrimination probability at 5$\,\sigma$ was found to be 60~Mt $\times$ year at 40 GeV.  This number can be significantly reduced by improving the detection efficiency in the 5--10~GeV muon energy region. Another way to increase the discrimination power, although with a weaker impact, relies on the lowering of  the detection energy threshold down to 1~GeV. However,  achieving the needed resolutions might be challenging below 5~GeV. A critical improvement could come from the reconstruction of the associated hadronic shower, not considered in this work. 

The discrimination power can also be  affected by the systematics on the neutrino production, oscillation, and interaction models. The effects of the uncertainties of the cross sections were not investigated in this study, since they are well constrained above 5~GeV. Uncertainties on the Earth density profile were shown to have a negligible impact on the NMH determination. The effect of differences in the predictions for the  energy and angular dependence of the neutrino flux, from the Honda, Bartol, and FLUKA models, was also demonstrated to be small. However, these models significantly differ in the overall flux normalization, which can result in a critical reduction of the discrimination power. This loss can be attenuated by anchoring the flux at high energies ($>20$~GeV), where the dependence on the NMH is small, or by removing  the dependence of the likelihood ratio test statistic on the flux normalization.
 
Finally, the impact of the oscillation parameter uncertainties and correlations on the discrimination power was carefully investigated. An important dependence of the NMH determination  on the values of $\theta_{13}$, $\theta_{23}$ and $\Delta m^2_{31}$, and  a weak one  on $\delta_{CP}$ and on  the shift between $\Delta m^2_{31}$(NH) and  $\Delta m^2_{31}$(IH), were observed. The effects induced by $\theta_{12}$ and $\Delta m^2_{12}$  were shown to be negligible.

 An extension of this work is foreseen in order to include  background events in the simulations, and more realistic detector resolution models, as soon as they will be available.

\acknowledgments
We are grateful to Mattias Blennow, Aart Heijboer, Eric Chassande--Mottin, Laurent Metivier, Elisa Resconi and Thomas Schwetz for very useful discussions.
We acknowledge the support of the UnivEarthS LabEx programme of Sorbonne Paris Cit\'e. 

\end{document}